\documentclass[useAMS,usenatbib,a4paper]{mn2e}
\usepackage{times}
\usepackage{amsmath}
\usepackage{amssymb}
\usepackage{graphicx}
\usepackage{xspace}
\usepackage{url}
\usepackage{hyperref}
\usepackage{color}
\usepackage[usenames,dvipsnames]{xcolor}
\usepackage[varg]{txfonts}
\usepackage[T1]{fontenc}
\usepackage[normalem]{ulem}
\def\tsys{$T_{\rm sys}$\xspace}

\def\path{./figures}

\usepackage{soul}

\newcommand{\aap}{A\&A}
\newcommand{\araa}{ARAA}
\newcommand{\mnras}{MNRAS}
\newcommand{\apjl}{ApJL}
\newcommand{\apjs}{ApJS}
\newcommand{\apj}{ApJ}
\newcommand{\aj}{ApJ}
\newcommand{\nat}{Nature}
\newcommand{\aapr}{A\&AR}
\newcommand{\pasp}{PASP}

\title[Radio variability of LS~I~+61~303]{Orbital and superorbital variability of LS~I~+61~303 at low radio frequencies with GMRT and LOFAR}

\author[B.~Marcote et al.]{B. Marcote$^{1,}$\thanks{E-mail: \url{bmarcote@am.ub.es} (BM); \url{mribo@am.ub.es} (MR); \url{jmparedes@ub.edu} (JMP)},
    M. Rib\'o$^{1,}$\thanks{Serra H\'unter Fellow.},
    J. M. Paredes$^{1}$,
    C. H. Ishwara-Chandra$^{2}$,
    J.~D.~Swinbank$^{3}$,\newauthor
    J.~W.~Broderick$^{4,5}$,
    S.~Markoff$^{6}$,
    R.~Fender$^{4}$,
    R.~A.~M.~J.~Wijers$^{6}$,
    G.~G.~Pooley$^{7}$,\newauthor
    A.~J.~Stewart$^{4}$,
    M.~E.~Bell$^{8,9}$,
    R.~P.~Breton$^{10,5}$,
    D.~Carbone$^6$,
    S.~Corbel$^{11,12}$,
    J.~Eisl\"offel$^{13}$,\newauthor
    H.~Falcke$^{14,15}$,
    J.-M.~Grie{\ss}meier$^{16,12}$,
    M.~Kuniyoshi$^{17}$,
    M.~Pietka$^{18}$,
    A.~Rowlinson$^{8}$,\newauthor
    M.~Serylak$^{19,20,12}$,
    A.~J.~van~der~Horst$^{21}$,
    J.~van~Leeuwen$^{15,6}$,
    M.~W.~Wise$^{15,6}$,
    P.~Zarka$^{22,12}$\\
$^{1}$Departament d'Astronomia i Meteorologia, Institut de Ci\`encies del Cosmos (ICCUB), Universitat de Barcelona (IEEC-UB), Mart\'{\i} i Franqu\`es 1,\\ E08028 Barcelona, Spain\\
$^{2}$National Centre for Radio Astrophysics, TIFR, Post Bag 3, Ganeshkhind, 411007, Pune, India\\
$^{3}$Department of Astrophysical Sciences, Princeton University, Princeton, NJ08544, USA\\
$^{4}$Astrophysics, Department of Physics, University of Oxford, Keble Road, Oxford OX1 3RH, UK\\
$^{5}$Department of Physics and Astronomy, University of Southampton, Highfield, Southampton, SO17 1BJ, UK\\
$^{6}$Anton Pannekoek Institute for Astronomy, University of Amsterdam, Science Park 904, 1098 XH Amsterdam, The Netherlands\\
$^{7}$Mullard Radio Astronomy Observatory, Cavendish Laboratory, The University of Cambridge, J.J. Thomson Avenue, Cambridge CB3~0HE, UK\\
$^{8}$CSIRO Astronomy and Space Science, PO Box 76, Epping, NSW 1710, Australia\\
$^{9}$ARC Centre of Excellence for All-sky Astrophysics (CAASTRO), The University of Sydney, NSW 2006, Australia\\
$^{10}$Jodrell Bank Centre for Astrophysics, School of Physics and Astronomy, The University of Manchester, Manchester M13 9PL, UK\\
$^{11}$Laboratoire AIM (CEA/IRFU - CNRS/INSU - Universit\'e Paris Diderot), CEA DSM/IRFU/SAp, F-91191 Gif-sur-Yvette, France\\
$^{12}$Station de Radioastronomie de Nan\c{c}ay, Observatoire de Paris, PSL Research University, CNRS, Univ. Orl\'{e}ans, OSUC, 18330 Nan\c{c}ay, France\\
$^{13}$Th\"uringer Landessternwarte, Sternwarte 5, D-07778 Tautenburg, Germany\\
$^{14}$Department of Astrophysics/IMAPP, Radboud University Nijmegen, PO Box 9010, 6500 GL Nijmegen, The Netherlands\\
$^{15}$ASTRON, the Netherlands Institute for Radio Astronomy, Postbus 2, 7990 AA, Dwingeloo, The Netherlands\\
$^{16}$LPC2E - Universit\'{e} d'Orl\'{e}ans / CNRS, 45071 Orl\'{e}ans cedex 2, France\\
$^{17}$NAOJ Chile Observatory, National Astronomical Observatory of Japan, 2-21-1 Osawa, Mitaka, Tokyo 181-8588, Japan\\
$^{18}$Oxford Astrophysics, Denys Wilkinson Building, Keble Road, Oxford OX1 3RH, UK\\
$^{19}$Department of Physics \& Astronomy, University of the Western Cape, Private Bag X17, Bellville 7535, South Africa\\
$^{20}$SKA South Africa, 3rd Floor, The Park, Park Road, Pinelands, 7405, South Africa\\
$^{21}$Department of Physics, The George Washington University, 725 21st Street NW, Washington, DC 20052, USA\\
$^{22}$LESIA, Observatoire de Paris, CNRS, UPMC, Universit\'e Paris-Diderot, 5 place Jules Janssen, 92195 Meudon, France
\vspace{-10pt}
}

\begin{document}

\date{}

\pagerange{\pageref{firstpage}--\pageref{lastpage}}
\pubyear{2015}

\maketitle

\label{firstpage}

\begin{abstract}
LS~I~+61~303 is a gamma-ray binary that exhibits an outburst at GHz frequencies each orbital cycle of $\approx$$26.5~\mathrm{d}$ and a superorbital modulation with a period of $\approx$4.6~yr. We have performed a detailed study of the low-frequency radio emission of LS~I~+61~303 by analysing all the archival GMRT data at 150, 235 and 610~MHz, and conducting regular LOFAR observations within the Radio Sky Monitor (RSM) at 150~MHz.
We have detected the source for the first time at 150~MHz, which is also the first detection of a gamma-ray binary at such a low frequency. We have obtained the light-curves of the source at 150, 235 and 610~MHz, all of them showing orbital modulation. The light-curves at 235 and 610~MHz also show the existence of superorbital variability.
A comparison with contemporaneous 15-GHz data shows remarkable differences with these light-curves.
At 15~GHz we see clear outbursts, whereas at low frequencies we see variability with wide maxima. The light-curve at 235~MHz seems to be anticorrelated with the one at 610~MHz, implying a shift of $\sim$0.5 orbital phases in the maxima. We model the shifts between the maxima at different frequencies as due to changes in the physical parameters of the emitting region assuming either free-free absorption or synchrotron self-absorption, obtaining expansion velocities for this region close to the stellar wind velocity with both mechanisms.
\end{abstract}

\begin{keywords}
radiation mechanisms: non-thermal -- binaries: close -- stars: individual: LS~I~+61~303 -- gamma rays: stars -- radio continuum: stars.\vspace{-30pt} 
\end{keywords}

\section{Introduction}\label{sec:intro}

Binary systems comprising a young massive star and a compact object can exhibit non-thermal emission from radio to gamma rays. A fraction of these systems displays a spectral energy distribution dominated by the MeV-GeV photons, thus becoming classified as gamma-ray binaries (see \citealt{dubus2013} for a review).
Only five gamma-ray binary systems are known at present: PSR~B1259$-$63 \citep{aharonian2005psr}, LS~5039 \citep{aharonian2005ls5039}, LS~I~+61~303 \citep{albert2006}, HESS~J0632+057 \citep{hinton2009,skilton2009}, and 1FGL~J1018.6$-$5856 \citep{fermi2012}. PSR~B1259$-$63 is the only system which is confirmed to host a pulsar; the natures of the compact objects are currently unknown for the remaining sources.

The multiwavelength emission from these sources can be explained by synchrotron and inverse Compton (IC) emission, within the context of various models. Although the particle acceleration mechanisms remain unclear, two different scenarios have been proposed. On the one hand, the microquasar scenario posits the existence of an accretion disc, corona and relativistic jets in the system, with the particle acceleration driven by accretion/ejection processes \citep{mirabel2006}. On the other hand, the young non-accreting pulsar scenario establishes the existence of a shock between the relativistic wind of the pulsar and the non-relativistic wind of the companion star, where particles can be accelerated up to relativistic energies \citep{dubus2013}.

The gamma-ray binary LS~I~+61~303 (also known as V615~Cas) is a system composed of a young main sequence B0~Ve star and a compact object orbiting it with a period $P_{\rm orb} = 26.4960 \pm 0.0028~\mathrm{d}$ \citep{gregory2002} and an eccentricity $e = 0.72 \pm 0.15$ \citep{casares2005lsi61303}. The system is located at $2.0 \pm 0.2~\mathrm{kpc}$ according to H~I measurements \citep{frail1991}, with coordinates $\alpha_{\rm J2000} = 2^{\rm h} 40^{\rm m} 31.7^{\rm s}$ and $\delta_{\rm J2000} = +61\degr 13' 45.6''$ \citep{moldon2012thesis}. Assuming the epoch of the first radio observation as the origin of the orbital phase, ${\rm JD_0} = 2\,443\,366.775$, the periastron passage takes place in the orbital phase range $0.23$--$0.28$ \citep{casares2005lsi61303, aragona2009}.
LS~I~+61~303 was initially identified as the counterpart of a gamma-ray source detected by {\em COS~B} \citep{hermsen1977}, and subsequently an X-ray counterpart was also detected \citep{bignami1981}. X-ray observations with {\em RXTE}/ASM revealed an orbital X-ray modulation \citep{paredes1997}. The system was also coincident with an EGRET source above 100~MeV \citep{kniffen1997} and finally it was detected as a TeV emitter by MAGIC \citep{albert2006}, which was later confirmed by VERITAS \citep{acciari2008}. An orbital modulation of the TeV emission was also found by MAGIC \citep{albert2009}. A correlation between the TeV and X-ray emission was revealed using MAGIC and X-ray observations, which together with the observed X-ray/TeV flux ratio, supports leptonic models with synchrotron X-ray emission and IC TeV emission (\citealt{anderhub2009}; but see \citealt{acciari2009}). Finally, {\em Fermi}/LAT reported GeV emission, orbitally modulated and anti-correlated with respect to the X-ray/TeV emission \citep{abdo2009}.

The 1--10~GHz radio light-curve of LS~I~+61~303 is also orbitally modulated, showing a clear outburst each orbital cycle that increases the $\la$\,50-mJy steady flux density emission up to $\sim$100--200~mJy. These outbursts are periodic, although changes in their shape and intensity have been reported from cycle to cycle (\citealt*{paredes1990}; \citealt{ray1997}). \citet{strickman1998} studied the evolution of a single outburst with multifrequency observations in the range 0.33--23~GHz. The outburst is detected at all frequencies, but the flux density peaks first at the highest frequencies, and later at the lower ones. A low-frequency turnover was also suggested to be present in the range 0.3--1.4~GHz \citep{strickman1998}.

A long-term modulation is also observed at all wavelengths in LS~I~+61~303, the so-called superorbital modulation, with a period $P_{\rm so} = 1\,667 \pm 8~\mathrm{d}$ or $\sim$4.6~yr \citep{gregory2002}. This modulation was firstly found at GHz radio frequencies \citep{paredes1987, gregory1989, paredes1990}, affecting the amplitude of the non-thermal outbursts and the orbital phases at which the onset and peak of these outbursts take place, drifting from orbital phases of $\sim$0.45 to $\sim$0.95 \citep{gregory2002}. The source exhibits the minimum activity at GHz frequencies during the superorbital phase range of $\phi_{\rm so} \sim 0.2$--$0.5$, whereas the maximum activity takes place at $\phi_{\rm so} \sim 0.78$--$0.05$, assuming the same ${\rm JD_0}$ as for the orbital phase. A similar behaviour is also observed in optical photometric and H$\alpha$ equivalent width observations that trace the thermal emission of the source \citep{paredes-fortuny2015}.
The origin of the superorbital modulation could be related to periodic changes in the circumstellar disc and the mass-loss rate of the Be star \citep{zamanov2013}, although other interpretations within the framework of a precessing jet are still discussed \citep[see][and references therein]{massi2014}.

At milliarcsecond scales LS~I~+61~303 has been observed and resolved several times \citep{massi2001, massi2004, dhawan2006, albert2008, moldon2012thesis}. \citet*{dhawan2006} showed a changing morphology as a function of the orbital phase, resembling a cometary tail.
\citet{albert2008} observed similar morphological structures at similar orbital phases to those considered in \citet{dhawan2006}. Later on, \citet{moldon2012thesis} also reported this behaviour: the morphology of LS~I~+61~303 changes periodically within the orbital phase. These morphological changes have been interpreted as evidence of the pulsar scenario \citep{dubus2006,dubus2013}, although other interpretations have also been suggested \citep*[see][and references therein]{massi2012}. It must be noted that radio pulsation searches have also been conducted with unsuccessful results \citep{mcswain2011,canellas2012}.

At the lowest frequencies ($\la 1~\mathrm{GHz}$), only a few observations have been published up to now. \citet{pandey2007} conducted two observations of LS~I~+61~303 at 235 and 610~MHz simultaneously with the Giant Metrewave Radio Telescope (GMRT: \citealt{ananthakrishnan1995}). They observed a positive spectral index of $\alpha \approx 1.3$ (defined as $S_{\nu} \propto \nu^{\alpha}$, where $S_{\nu}$ is the flux density at a frequency $\nu$) in both epochs and reported variability at both frequencies at a 2.5 and 20-$\sigma$ level, respectively. Although \citet{strickman1998} also observed the source at 330~MHz three times during one outburst, given the large uncertainties of the measured flux densities they could not infer variability at more than the $1$-$\sigma$ level.
At these low frequencies we should observe the presence of different absorption mechanisms, such as synchrotron self-absorption (SSA), free-free absorption (FFA) or the Razin effect, as found for other gamma-ray binaries (see e.g. \citealt{marcote2015ls5039} for the case of LS~5039). In addition, at these low frequencies we should also expect extended emission at about arcsec scales originating from the synchrotron emission from low-energy particles \citep{bosch-ramon2009ls5039,durant2011}.

In this paper we present the first deep and detailed study of the radio emission from LS~I~+61~303 at low frequencies through GMRT and LOw Frequency ARray \citep[LOFAR:][]{vanhaarlem2013} data to unveil its behaviour along the orbit. We compare the results with contemporaneous high-frequency observations conducted with the Ryle Telescope (RT) and the Owens Valley Radio Observatory (OVRO) 40-m telescope. In Sect.~\ref{sec:observations} we present all the radio data analysed in this work together with the data reduction and analysis processes. In Sect.~\ref{sec:results} we present the obtained results and in Sect.~\ref{sec:discussion} we discuss the observed behaviour of LS~I~+61~303 in the context of the known orbital and superorbital variability. Our conclusions are presented in Sect.~\ref{sec:conclusions}.

\section{Observations and data reduction}\label{sec:observations}

\begin{figure}
    \includegraphics[width=0.475\textwidth]{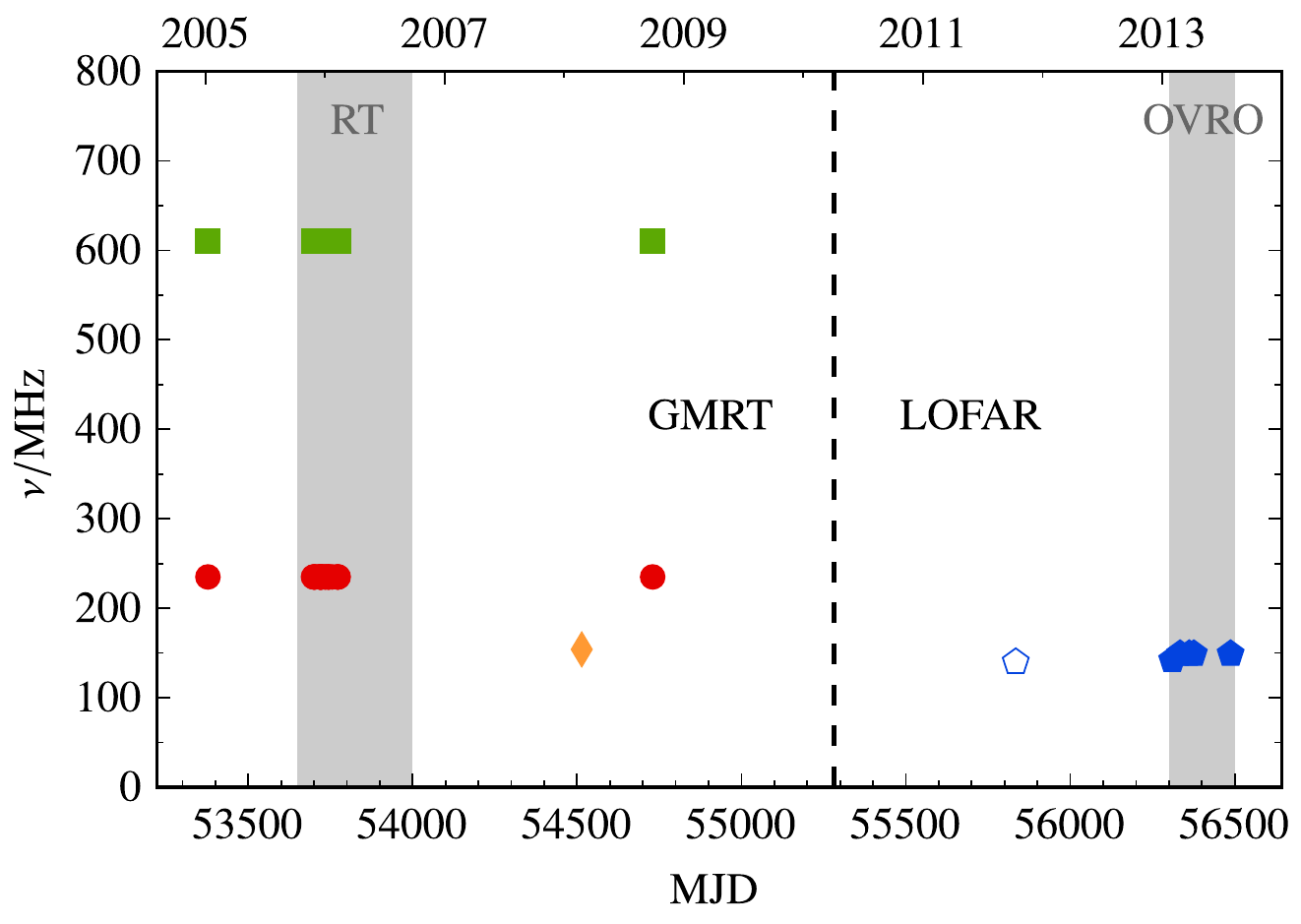}
    \caption{Summary of all the data presented in this work in a frequency versus time diagram (Modified Julian Date, MJD, on the bottom axis and calendar year on the top axis). The vertical dashed line separates the GMRT data (left) from the LOFAR data (right). Green squares represent the 610-MHz GMRT observations, red circles correspond to the simultaneous GMRT observations at 235~MHz, the orange diamond shows the 154-MHz GMRT data and the blue pentagons represent the LOFAR observations at $\approx$150~MHz. The open pentagon corresponds to the commissioning LOFAR observation, which is not considered in the results of this work. The grey bands indicate the time intervals during which 15-GHz RT and OVRO observations used in this work were conducted.}
    \label{fig:summary-obs}
\end{figure}

To reveal the orbital behaviour of LS~I~+61~303 at low radio frequencies we have analysed data from different instruments and different epochs. These data include archival GMRT observations from a monitoring programme in 2005--2006, as well as three additional archival observations from 2005 and 2008. They also include a LOFAR commissioning observation in 2011 and five LOFAR observations performed in 2013. RT and OVRO observations at 15~GHz, contemporaneous to these low-frequency observations, have also been analysed to obtain a complete picture of the behaviour of the source.
In Fig.~\ref{fig:summary-obs} we summarize all these observations in a frequency versus time diagram. The first seven columns of Table~\ref{tab:data} shows the log of the low-frequency observations.

\begin{table*}
\def\s{~~~~~~}
\caption{Summary of the low-frequency data presented in this work. Each row corresponds to one observation for which we show the facility used, the project code of the observation, the date (in calendar date and Modified Julian Date, MJD) of the centre of the observation, the corresponding orbital phase $\phi_{\rm orb}$ (using $P_{\rm orb} = 26.4960~\mathrm{d}$ and ${\rm JD_0} = 2\,443\,366.775$), the corresponding superorbital phase (assuming $P_{\rm so} = 1\,667~\mathrm{d}$ and the same ${\rm JD_0}$), the total time on source ($t$), and the flux densities at each frequency with the 1-$\sigma$ uncertainty (3-$\sigma$ upper-limits in case of non-detections).}
\label{tab:data}
\begin{tabular}{l@{\s}c@{\s}c@{\s}c@{\s}c@{\s}c@{\s}c@{\s}r@{\;}c@{\;}l@{\s}r@{\;}c@{\;}l@{\s}r@{\;}c@{\;}l}
\hline
Facility & Project Code & Date & MJD & $\phi_{\rm orb}$ & $\phi_{\rm so}$ & $t$ &
\multicolumn{3}{c}{$S_{\approx150\,\mathrm{MHz}}$} &
\multicolumn{3}{c}{$S_{235\,\mathrm{MHz}}$} &
\multicolumn{3}{c}{$S_{610\,\mathrm{MHz}}$}\\
&&&&&& $(\mathrm{min})$&
\multicolumn{3}{c}{$(\mathrm{mJy})$} &
\multicolumn{3}{c}{$(\mathrm{mJy})$} &
\multicolumn{3}{c}{$(\mathrm{mJy})$}\\
\hline
 GMRT & \hphantom{4}\ 07PDA01$^{(1)}$ & 07/01/2005 & 53377.47 & 0.84& 0.01&30& &-& & 34 &$\pm$& 10 & 169 &$\pm$& 3\\
GMRT & 09PDA01 & 24/11/2005 & 53698.63 & 0.96 & 0.20& 82& &-& & \multicolumn{3}{c}{$<31$} & 89.9 &$\pm$& 1.3\\
GMRT & 09PDA01 & 25/11/2005 & 53699.71 & 0.00 & 0.20& 56& &-& & \multicolumn{3}{c}{$<30$} & 90 &$\pm$& 2\\
GMRT & 09PDA01 & 26/11/2005 & 53700.58 & 0.03 & 0.20& 82& &-& & 23 &$\pm$& 4 & 84.5 &$\pm$& 1.9\\
GMRT & 09PDA01 & 27/11/2005 & 53701.59 & 0.07 & 0.20& 164& &-& & 35 &$\pm$& 4 & 79.3 &$\pm$& 1.2\\
GMRT & 09PDA01 & 29/11/2005 & 53703.62 & 0.15 & 0.20& 80& &-& & 41 &$\pm$& 4 & 72.8 &$\pm$& 1.5\\
GMRT & 09PDA01 & 12/12/2005 & 53716.70 & 0.64 & 0.21& 60& &-& & 56 &$\pm$& 6 & 69.3 &$\pm$& 1.6\\
GMRT & 09PDA01 & 13/12/2005 & 53717.65 & 0.68 & 0.21& 70& &-& & 41 &$\pm$& 2 & 58.5 &$\pm$& 1.3\\
GMRT & 09PDA01 & 15/12/2005 & 53719.61 & 0.75 & 0.21& 80& &-& & 38 &$\pm$& 9 & 67.5 &$\pm$& 1.5\\
GMRT & 09PDA01 & 16/12/2005 & 53720.53 & 0.79 & 0.21& 524& &-& & 39.0 &$\pm$& 1.9 & 61.2 &$\pm$& 0.8\\
GMRT & 09PDA01 & 17/12/2005 & 53721.56 & 0.82 & 0.21& 130& &-& & 33 &$\pm$& 3 & 97.1 &$\pm$& 1.1\\
GMRT & 09PDA01 & 18/12/2005 & 53722.53 & 0.86 & 0.21& 216& &-& & 40 &$\pm$& 4 & 92.1 &$\pm$& 1.1\\
GMRT & 09PDA01 & 19/12/2005 & 53723.53 & 0.90 & 0.21& 80& &-& & 43 &$\pm$& 5 & 87.8 &$\pm$& 1.7\\
GMRT & 09PDA01 & 29/12/2005 & 53733.61 & 0.28 & 0.22& 80& &-& & 48 &$\pm$& 11 & 64.8 &$\pm$& 1.7\\
GMRT & 09PDA01 & 30/12/2005 & 53734.49 & 0.31 & 0.22& 80& &-& & &-& & 58.8 &$\pm$& 1.9\\
GMRT & 09PDA01 & 31/12/2005 & 53735.41 & 0.35 & 0.22& 56& &-& & 53 &$\pm$& 5 & 54 &$\pm$& 3\\
GMRT & 09PDA01 & 08/01/2006 & 53743.71 & 0.66 & 0.23& 41& &-& & \multicolumn{3}{c}{$<69$} & 26 &$\pm$& 3\\
GMRT & 09PDA01 & 09/01/2006 & 53744.36 & 0.68 & 0.23& 144& &-& & \multicolumn{3}{c}{$<46$} & 26.7 &$\pm$& 0.7\\
GMRT & 09PDA01 & 10/01/2006 & 53745.55 & 0.73 & 0.23& 82& &-& & 20 &$\pm$& 4 & 33.9 &$\pm$& 1.0\\
GMRT & 09PDA01 & 19/01/2006 & 53754.67 & 0.07 & 0.23& 82& &-& & &-& & 93.6 &$\pm$& 1.6\\
GMRT & 09PDA01 & 20/01/2006 & 53755.49 & 0.10 & 0.23& 70& &-& & 36 &$\pm$& 2 & 99.9 &$\pm$& 1.6\\
GMRT & 09PDA01 & 21/01/2006 & 53756.31 & 0.14 & 0.23& 48& &-& & &-& & 88.4 &$\pm$& 1.1\\
GMRT & 09PDA01 & 05/02/2006 & 53771.31 & 0.70 & 0.24& 60& &-& & 29 &$\pm$& 5 & 64 &$\pm$& 2\\
GMRT & 09PDA01 & 06/02/2006 & 53772.27 & 0.74 & 0.24& 82& &-& & \multicolumn{3}{c}{$<33$} & 75.9 &$\pm$& 1.4\\
GMRT & 09PDA01 & 07/02/2006 & 53773.61 & 0.79 & 0.24& 82& &-& & 22 &$\pm$& 5 & 90.3 &$\pm$& 1.7\\
GMRT & 13MPA01 & 18/02/2008 & 54514.28 & 0.74 & 0.69& 662& 52 &$\pm$& 11$^*$ & &-& & &-&\\
GMRT & 08DT051 & 20/09/2008 & 54729.90 & 0.88 & 0.82& 492& &-& & 140 &$\pm$& 2 & 337 &$\pm$& 3\\
LOFAR & L84298--L84317& 17/01/2013 & 56309.83 & 0.51 & 0.76&200 & 32 &$\pm$& 6$^\dagger$ & &-& & &-&\\
LOFAR & L89566--L89589& 10/02/2013 & 56333.71 & 0.41 & 0.78& 116& \multicolumn{3}{l}{$<24^{\dagger\dagger}$} & &-& & &-&\\
LOFAR & L100374--L100397& 10/03/2013 & 56361.67 & 0.47 & 0.80&116 & \multicolumn{3}{l}{$<28^{\dagger\dagger}$} & &-& & &-&\\
LOFAR & L107793--L107816& 24/03/2013 & 56375.58 & 0.99 & 0.80&116 & 77 &$\pm$& 9$^{\dagger\dagger}$ & &-& & &-&\\
LOFAR & L160534--L160557& 14/07/2013 & 56487.42 & 0.21 & 0.87&116 & 31 &$\pm$& 6$^{\dagger\dagger}$ & &-& & &-&\\

\hline
\end{tabular}
\flushleft {(1)~\citet{pandey2007}. $^*$ This observation was centred at 154~MHz with a bandwidth of 16~MHz. $^\dagger$ This observation was centred at 142~MHz with a bandwidth of 2.3~MHz. $^{\dagger\dagger}$ These observations were centred at 149~MHz with a bandwidth of 0.8~MHz.}
\end{table*}

\subsection{Archival GMRT observations}

The GMRT monitoring covers the observations performed between 2005 November 24 and 2006 February 7, at 235 and 610~MHz, simultaneously. These data sets include 25 observations spread over this time interval, all of them obtained with a single IF, with single circular polarisation at 235~MHz (LL) and 610~MHz (RR). The 235 and 610~MHz data have bandwidths of 8 and 16~MHz, divided into 64 and 128 channels, respectively. These observations display a wide range of observing times, from 30~min to 11~h (see column 7 in Table~\ref{tab:data}). 3C~48, 3C~147 and 3C~286 were used as amplitude calibrators, and 3C~119 (0432$+$416) or 3C~48 as phase calibrators.

In addition to the previous monitoring, there are also three isolated archival GMRT observations. Two of these observations were carried out on 2005 January 7 and on 2008 September 20, at 235 and 610~MHz, using the same setup as that described above. The remaining observation was performed at 154~MHz on 2008 February 18, with dual circular polarization (LL and RR) and a bandwidth of 16~MHz divided into 128 channels. 3C~48, 3C~147 and 3C~286 were again used as amplitude calibrators, and 3C~119 was the phase calibrator for the three observations.

The GMRT data were calibrated and analysed using standard procedures within {\sc aips}\footnote{The NRAO Astronomical Image Processing System, {\tt aips}, is maintained by the National Radio Astronomy Observatory (NRAO). \url{http://www.aips.nrao.edu}}, although {\sc obit}\footnote{\url{http://www.cv.nrao.edu/~bcotton/Obit.html}} \citep{cotton2008}, {\sc parseltongue}\footnote{\url{http://www.jive.nl/jivewiki/doku.php?id=parseltongue:parseltongue}} \citep{kettenis2006} and {\sc spam}\footnote{\url{https://safe.nrao.edu/wiki/bin/view/Main/HuibIntemaSpam}} \citep{intema2009} have also been used to develop scripts that call {\sc aips} tasks to easily reduce the whole data set.

We loaded and flagged the raw visibilities in {\sc aips}, removing telescope off-source times, instrumental problems or radio-frequency interference (RFI). A first calibration using a unique channel, free of RFI, was performed on the data, and after that we bandpass calibrated them. A more accurate flagging process was performed later on and we finally calibrated the full data set in amplitude and phase, taking into account the previous bandpass calibration.
The target source was imaged and self-calibrated several times to correct for phase and amplitude errors that were not properly corrected during the previous calibration, using a Briggs robust weighting of zero in the cleaning process \citep{briggs1995}. As the GMRT field of view at these low frequencies covers a few degrees (between $\sim$5 and $1.5\degr$ at 154 and 610~MHz, respectively), we needed to consider the full $uvw$-space during the imaging process \citep{thompson1999} and correct the final images for the primary beam attenuation. The GMRT observation performed on 2005 November 28 could not be properly calibrated and hence could not provide results, leading to a total of 24 flux density measurements. In the observations performed on 2005 December 30, 2006 January 19 and 21, the 235~MHz data could not be properly calibrated: the runs were relatively short and a large amount of data had to be flagged, thus providing clean images that were not good enough to self-calibrate the data, preventing the collection of reliable flux densities.

After this data reduction process, we also performed a correction of the system temperature, \tsys, for each antenna to subtract the contribution of the Galactic diffuse emission, relevant at low frequencies \citep[see a detailed explanation in the appendix of][]{marcote2015ls5039}. The obtained \tsys corrections were directly applied to the flux densities of the final target images. We note that these corrections imply an additional source of uncertainty that does not affect the relative flux density variations from one epoch to another at a given frequency. The uncertainties introduced in the flux densities are close to the typical rms values obtained at 235 and 610~MHz, but at 154~MHz the uncertainties increased from $\approx$5\% to $\approx$20\%.

The measurement of the flux densities in each image was done using the {\tt tvstat} and {\tt jmfit} tasks of {\sc aips}, both of them providing consistent values in all cases. We determined the root-mean-square (rms) noise of the images using {\tt tvstat} in a region around the source without including it nor any background source.

To guarantee the reliability of the measured flux densities we monitored the flux density values of four background sources detected in the field of view of LS~I~+61~303. The lack of a similar trend in all sources allows us to be confident that the variability described below is intrinsically related to LS~I~+61~303 and not due to calibration issues.

\subsection{LOFAR observations}

\begin{figure*}
    \includegraphics[height=0.44\textheight]{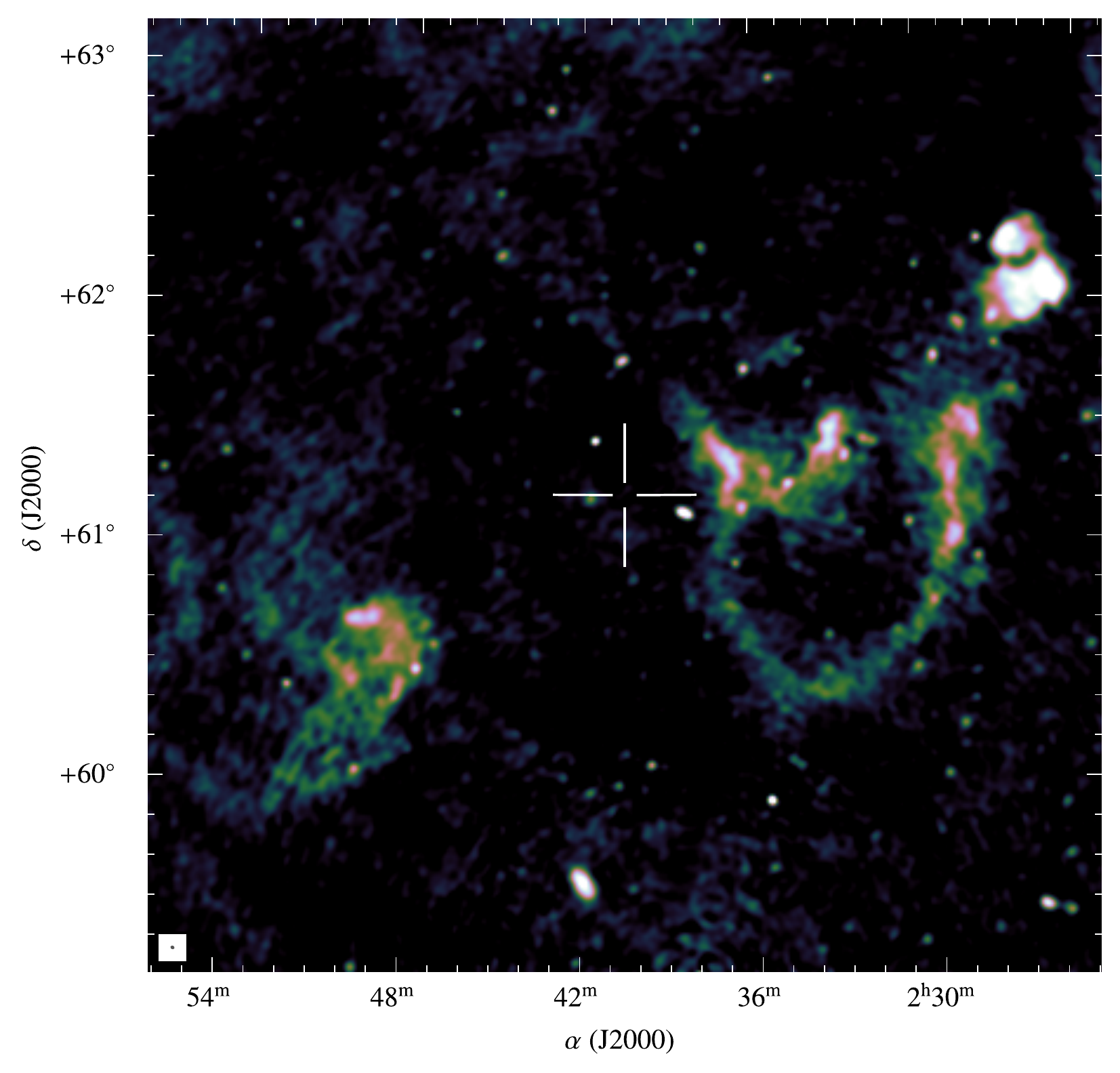}
    \includegraphics[height=0.43\textheight]{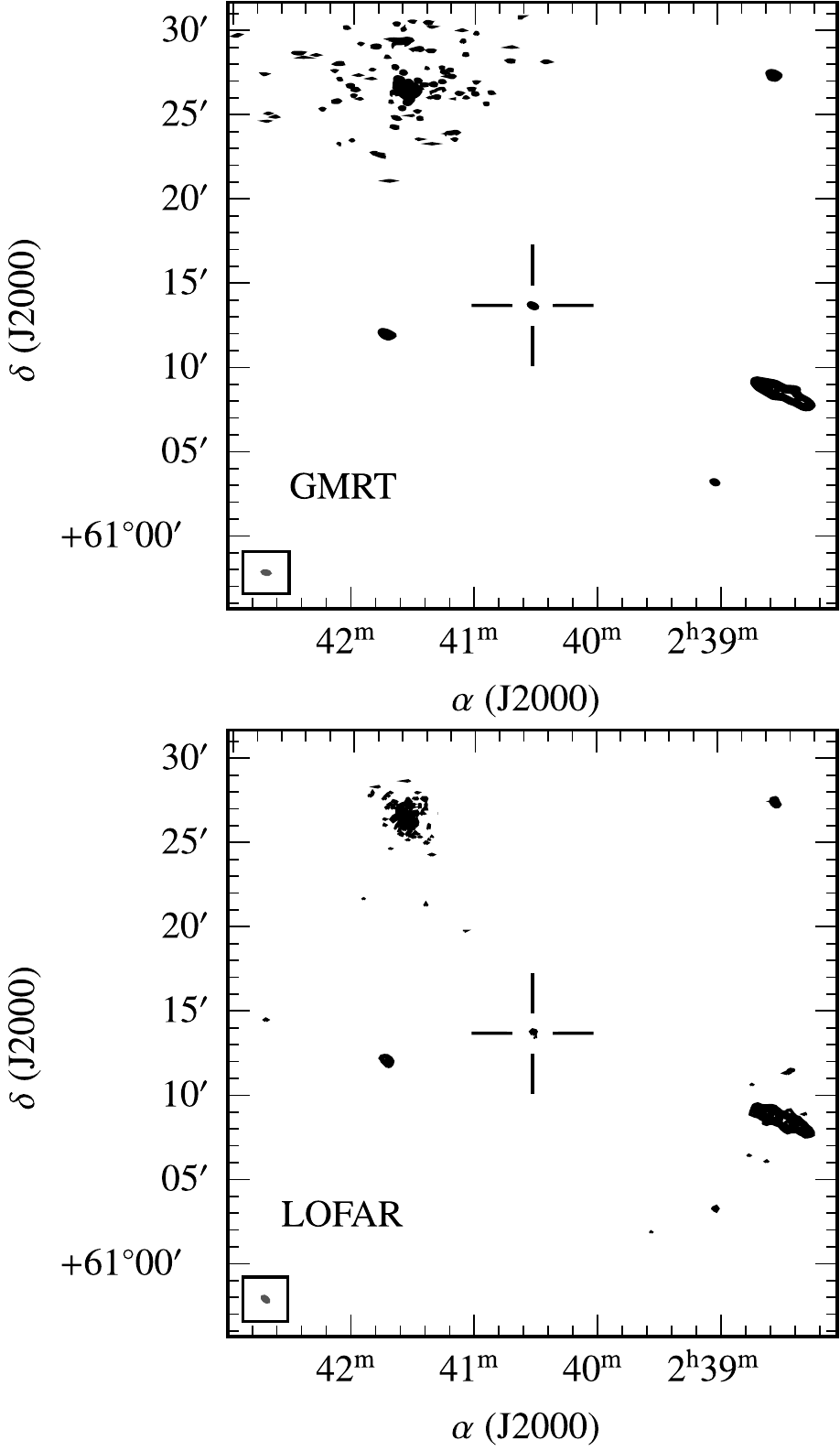}
    \caption{{\em Left}: field of LS~I~+61~303 observed with LOFAR at 149~MHz on 2013 July 14 without performing any cut in the $uv$-distance. The synthesized beam, shown in the lower left corner, is $23 \times 15~\mathrm{arcsec^2}$ with a PA of $64\degr$, and the rms is $8~\mathrm{mJy\ beam^{-1}}$ around the target source position. The positions of the sources present displacements of about half a synthesized beam, on average, with respect to the previously-recorded ones (see text). The location of LS~I~+61~303 is marked with the white cross. The source is masked by extended emission from the Heart Nebula in this image. {\em Right:} zoom around the position of LS~I~+61~303 in the 154-MHz GMRT observation (top) and in the 2013 March 24 149-MHz LOFAR observation applying the final cut at baselines below $0.2~\mathrm{k\lambda}$ (bottom). The source is clearly detected in both images. The synthesized beams, shown on the bottom left corner, are $27 \times 12~\mathrm{arcsec^2}$ with a PA of $80\degr$, and $27 \times 15~\mathrm{arcsec^2}$ with a PA $= 49\degr$, and the rms values are $11$ and $6~\mathrm{mJy\ beam^{-1}}$, respectively. Contours start at 5-$\sigma$ noise level and increase by factors of $2^{1/2}$. In the GMRT image (top), we have applied a shift of 25~arcsec to recover the proper astrometric positions. No shift is required in the LOFAR image (bottom).}
    \label{fig:image}
\end{figure*}
We have also analysed several LOFAR observations conducted with the High Band Antennas (HBAs). First, we conducted a deep 6-h LOFAR observation at 140~MHz during its commissioning stage on 2011 September 30 to test the system and estimate the expected behaviour of LS~I~+61~303 at these low frequencies. This observation was performed using 23 core stations plus 9 remote stations, with a total bandwidth of 48~MHz divided in 244 subbands.
These data show a large rms noise level, with possible large uncertainties in the absolute flux density scale related to the calibration process. Hence we will not use these in this work (see preliminary results of this observation in \citealt{marcote2012}).

In 2013, we conducted five 3-hr LOFAR observations at around 150~MHz within the Radio Sky Monitor (RSM), which observes regularly several variable or transient radio sources \citep{fender2008}. The target source was always centred on transit, using 23 core stations plus 13 remote stations. In four of these observations we observed the source at 149~MHz with a total bandwidth of $\approx$$0.8~\mathrm{MHz}$ divided into 4 subbands. 3C~48 or 3C~147 were used as amplitude calibrators, observed in runs of 2 min interleaved in 11-min on-source runs. The observation on 2013 January 17 was conducted with a different setup: 20-min on-source runs centred at 142~MHz and divided into 12 subbands, with a total bandwidth of 2.4~MHz.

The LOFAR data were calibrated using standard procedures within the LOFAR Imaging Tools ({\sc lofim}, version 2.5.2, see \citealt{heald2010} and \citealt{vanhaarlem2013} for a detailed explanation).
The data were initially flagged using standard settings of the {\sc aoflagger} \citep*{offringa2012}, and averaged in time and frequency with the LOFAR New Default Pre-Processing Pipeline ({\sc ndppp}) with an integration time of 10~s and 4 channels per subband.
In general, very bright sources did not need to be demixed from the target data sets, as they did not contribute significantly to the visibilities\footnote{The demixing process consists of removing from the target field visibilities the interference produced by the strongest off-axis sources in the sky, the so-called A-team: Cyg~A, Cas~A, Tau~A, Vir~A, Her~A and Hyd~A.}.

The calibration was performed on each subband individually with the BlackBoard Selfcal ({\sc bbs}) package using standard settings, and the solutions were transferred to the target field. We used an initial sky model for the field of LS~I~+61~303 based on the VLA Low-Frequency Sky Survey (VLSS, \citealt{cohen2007}).
A manual flagging was performed using the {\tt casaviewer} and {\tt casaplotms} tasks from {\sc casa}\footnote{The Common Astronomy Software Applications, {\sc casa}, is developed by the National Radio Astronomy Observatory (NRAO). \url{http://casa.nrao.edu}} and the imaging process was conducted with a development version of {\sc awimager} \citep{tasse2013}. The resulting model was used in a later phase self-calibration, and we imaged the data again. The last two tasks were performed recursively between 2 and 5 times until the solutions converged.
Due to the bright extended emission that is detected in the field of LS~I~+61~303 (located in the region of the Heart Nebula, see Fig.~\ref{fig:image}) we only kept baselines larger than $0.2~\mathrm{k\lambda}$ (or 400~m) in subsequent analyses. Although the maximum baseline in the recorded data was $\sim$$85~\mathrm{k\lambda}$ (or 170~km), the baselines $\ga$$8~\mathrm{k\lambda}$ (17~km) were removed during the data reduction (either during the manual flagging or during the self-calibration process). We used a Briggs robust weighting of zero again, because it produced the lowest noise level in the final images, and a pixel size of 10~arcsec.

We used the {\tt imstat} and {\tt imfit} tasks from {\sc casa} (both equivalent to the {\tt tvstat} and {\tt jmfit} tasks from {\sc aips} used for the GMRT data) to measure the LS~I~+61~303 flux densities and the rms of the images.
To guarantee the reliability of the measured flux densities in the LOFAR images, we monitored the same four background sources chosen in the GMRT images. The lack of a similar trend in all sources, exhibiting consistent values with the ones inferred from the GMRT images, allows us again to be confident about the absence of flux calibration issues above the noise level in the LOFAR data.

We note that given the large field of view of these low-frequency images (either from the GMRT data or from the LOFAR ones), we detect differences in the ionospheric refractive effects for different regions across the field of view. These effects, in combination with the self-calibration cycles performed during the reduction process, produce slight displacements of the sources from their original positions. In general, it is common to end up with differences of up to the order of the synthesized beam size, although with differences in moduli and direction for different regions of the field.

\subsection{Complementary 15-GHz observations}

Two complementary observing campaigns that were conducted at the same epochs as the GMRT monitoring and the LOFAR observations have also been included in this work.

The Ryle Telescope \citep{pooley1997} has observed LS~I~+61~303 at 15~GHz over many years, with contemporaneous observations during the epoch at which the 2005--2006 GMRT monitoring was performed. The observations are centred at 15.2~GHz, recording Stokes $I+Q$ and a bandwidth of 350~MHz. The four mobile and one fixed antennas were arranged in a compact configuration, with a maximum baseline of 100~m.

The observing technique is similar to that described in \citet{pooley1997}. Given that we used baselines up to 100~m, we obtained a resolution in mapping mode of about 30~arcsec. The observations included regular visits to a phase calibrator (we used J0228+6721) to allow corrections for slow drifts of instrumental phase; the flux-density scale was established by nearby observations of 3C~48, 3C~147 or 3C~286.

Observations of LS~I~+61~303 were also conducted with the OVRO 40-m dish covering the epoch of the 2013 LOFAR observations. These data are part of a long-term monitoring that has been presented in \citet*{massi2015}. The data were reduced by these authors following the procedures described in \citet{richards2011}.

\section{Results}\label{sec:results}

LS~I~+61~303 appears as a point-like source in all the images.
The resulting synthesized beams for the GMRT data range from $30 \times 14~\mathrm{arcsec^2}$ to $15 \times 5.4~\mathrm{arcsec^2}$, from 154 to 610~MHz, respectively. The synthesized beam for the LOFAR data is about $20 \times 15~\mathrm{arcsec^2}$.
The left panel of Fig.~\ref{fig:image} shows the field of LS~I~+61~303 obtained from the LOFAR observation conducted on 2013 July 14 without applying any $uv$-cut during the imaging process. The zoomed images (Fig.~\ref{fig:image}, right) show the source as seen from the 154-MHz GMRT observation and from one of the LOFAR runs that we have analyzed in this work.
The flux density values of all the GMRT and LOFAR observations are shown in Table~\ref{tab:data}. In this section we discuss the light-curves obtained from the 235/610-MHz GMRT monitoring and from the $\approx$150-MHz LOFAR observations, including the 154-MHz GMRT data.

\subsection{Light-curves at 235 and 610~MHz}

\begin{figure}
    \includegraphics[width=0.483\textwidth]{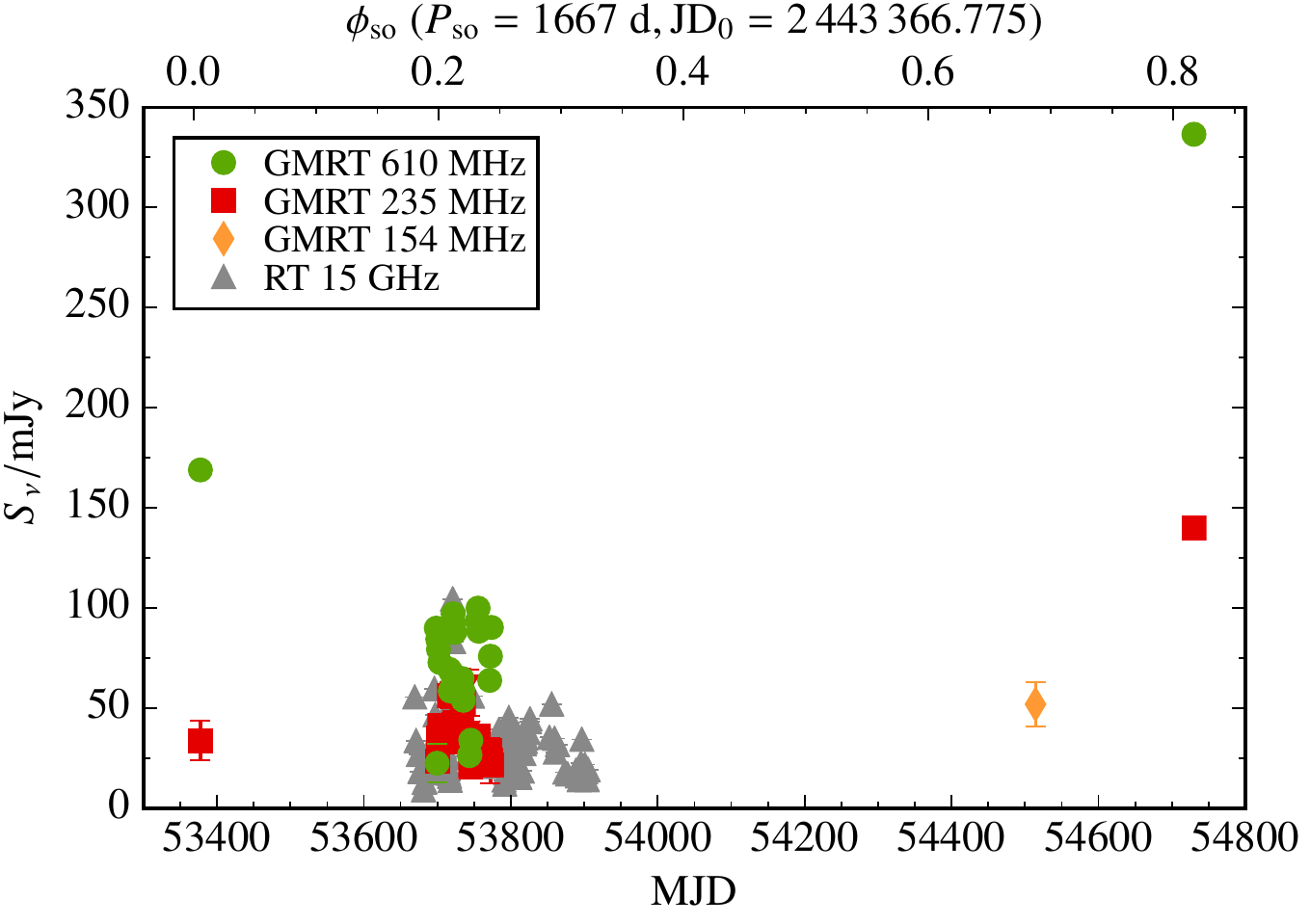}
    \caption{Flux density values of LS~I~+61~303 as a function of the MJD obtained from all the analysed GMRT data (2005--2008). We show the superorbital phase, $\phi_{\rm so}$, on the top $x$-axis. The green circles represent the 610-MHz data, the red squares the 235-MHz data, and the orange diamond the 154-MHz ones. The grey triangles represent the daily averages of the RT data at 15~GHz. Error bars represent 1-$\sigma$ uncertainties (smaller than the data points when they are not visible). We note that the source exhibits larger flux densities during the isolated GMRT observations (at superorbital phases of $\phi_{\rm so}\approx 0.80$--$0.95$) than during the GMRT monitoring ($\phi_{\rm so} \approx 0.2$).}
    \label{fig:gmrt-mjd-all}
\end{figure}
The dual 235/610-MHz mode in the GMRT observations allows us to observe the state of LS~I+61~303 and its evolution simultaneously at these two frequencies. Fig.~\ref{fig:gmrt-mjd-all} shows the flux densities of all the analysed GMRT observations (the GMRT monitoring at 235 and 610~MHz and the three isolated GMRT observations) as a function of the Modified Julian Date (MJD, bottom axis) and the superorbital phase ($\phi_{\rm so}$, top axis).
We observe that the flux densities obtained in the GMRT monitoring at 610~MHz (conducted at $\phi_{\rm so} \approx 0.2$, around the minimum activity of LS~I~+61~303 at GHz frequencies) are significantly lower than the ones obtained in the other two observations (conducted at $\phi_{\rm so} \approx 0.8$--$0.0$, during the maximum activity). We also observe this effect at 235~MHz with respect to the observation at $\phi_{\rm so} \approx 0.8$ (but not at $\phi_{\rm so} \approx 0.0$). This indicates the existence of the superorbital variability at frequencies below 1~GHz.

\begin{figure}
    \includegraphics[width=0.483\textwidth]{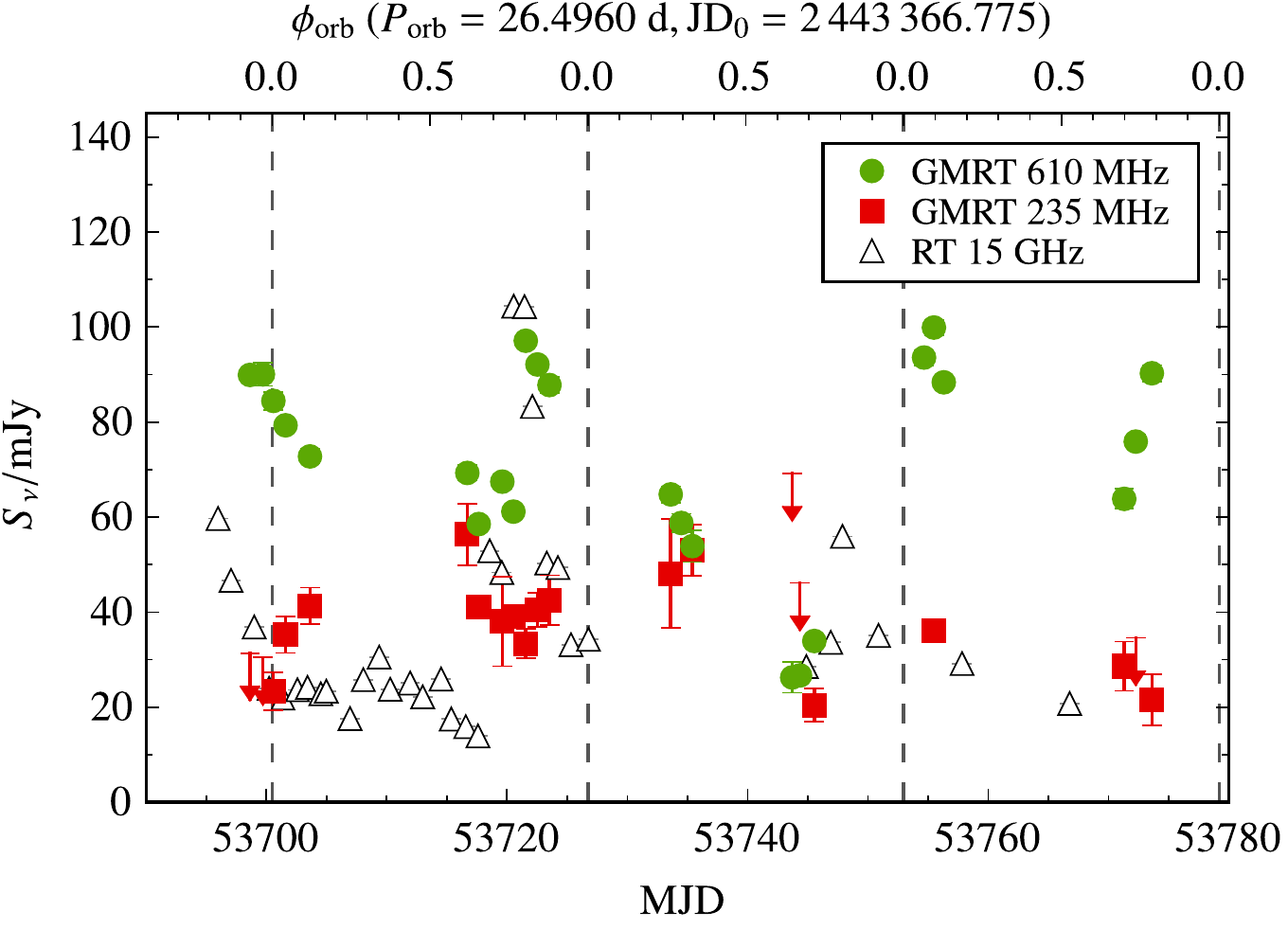}
    \caption{Same as Fig.~\ref{fig:gmrt-mjd-all} but zooming in on the GMRT monitoring observations. In this case we show the orbital phase, $\phi_{\rm orb}$, on the top $x$-axis. The vertical dashed lines show the epochs at which $\phi_{\rm orb} = 0$. We observe the presence of enhanced emission at 610~MHz coincident with the outbursts at 15~GHz but with a slower decay (see MJD~52695--52705, 53720--53725). In contrast, at 235~MHz we observe a smaller degree of variability, with no clear orbital trends. The arrows represent the 3-$\sigma$ upper-limits.}
    \label{fig:gmrt-mjd}
\end{figure}
Fig.~\ref{fig:gmrt-mjd} focuses on the GMRT monitoring, showing the 610 and 235-MHz light-curve of LS~I~+61~303 along three consecutive orbital cycles. The 610-MHz data exhibit variability with maxima roughly coincident with the outbursts observed in the 15-GHz RT data (e.g. see $\mathrm{MJD} \sim 53720$). However, at 610~MHz we observe that the decay of the emission is slower (e.g. compare the decays in MJD~53695--53705, 53720--53725). The average flux density during the whole monitoring is $70~\mathrm{mJy}$, with a standard deviation of $20~\mathrm{mJy}$ and a significant variability. For these variability analyses (and in the rest of this work) we have taken the most conservative choice to consider the 3-$\sigma$ upper-limits as the possible flux density value of the source, assuming $S_{\nu} \approx 3\sigma \pm \sigma$.
At 235~MHz we observe a smaller degree of variability, with no clear correlation with the previous one. In this case we infer an average flux density of $37 \pm 8~\mathrm{mJy}$ and a variability at a 6-$\sigma$ confidence level.

\begin{figure}
    \includegraphics[width=0.475\textwidth]{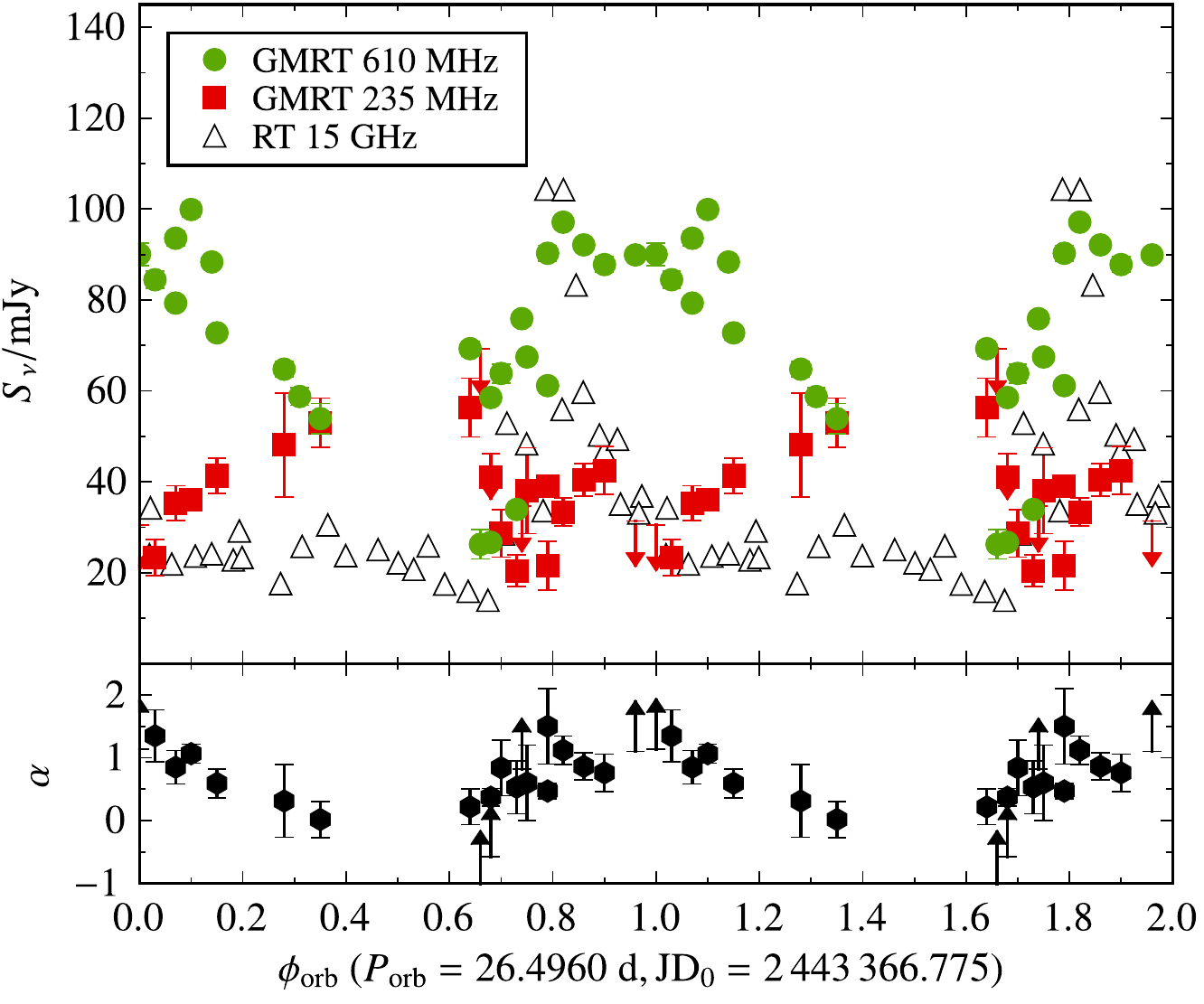}
    \caption{{\em Top:} folded light-curve with the orbital period of the data shown in Fig.~\ref{fig:gmrt-mjd}. The 610~MHz data show a quasi-sinusoidal modulation with enhanced emission at $\phi_{\rm orb} \approx 0.8$--$1.1$, whereas the 15-GHz outbursts take place at $\phi_{\rm orb} \approx 0.8$ with a fast decay. The 235~MHz light-curve is almost anticorrelated with the one observed at 610~MHz. 
    {\em Bottom:} spectral index $\alpha$ derived from the GMRT data presented above. Open hexagons represent the spectral index from the isolated GMRT data (not shown on top). Error bars represent 1-$\sigma$ uncertainties and the arrows represent the 3-$\sigma$ upper/lower-limits.}
    \label{fig:gmrt-phase}
\end{figure}

Folding these data with the orbital phase (Fig.~\ref{fig:gmrt-phase}, top), we clearly see that the radio emission is orbitally modulated.
At 610~MHz we observe that the enhanced emission takes place in the range of $\phi_{\rm orb} \sim 0.8$--$1.1$, whereas at 15~GHz the maximum occurs at $\phi_{\rm orb} \approx 0.8$ and shows a faster decay.
At 235-MHz we observe that the flux density increases between $\phi_{\rm orb} \sim 0.0$ and $0.4$. This phase range is followed by an interval without data between phases 0.4 and 0.6, after which we observe the largest flux density value followed by a fast decrease in the flux density. The increase and decrease of the flux density are traced by 4--5 data points in each case. Therefore, we observe that the maximum emission probably occurs in the range between $\phi_{\rm orb} \approx 0.3$ and $\approx0.7$, which is the location of the minimum at 610~MHz. In fact, the 235~MHz and the 610~MHz light-curves are almost anticorrelated. 
Fig.~\ref{fig:gmrt-phase} (bottom) shows the spectral index $\alpha$ (determined from the 235 and 610~MHz data) as a function of the orbital phase. The spectral index is also orbitally modulated, following essentially the 610-MHz flux density emission.
From the isolated GMRT observations we obtain a spectral index coincident with the modulation observed from the GMRT monitoring, despite the much larger flux density values (see open hexagons in Fig.~\ref{fig:gmrt-phase}, bottom).
We note that Fig.~\ref{fig:gmrt-phase} shows data from different orbital cycles that have been poorly sampled. In addition, we remark that the outbursts seen at GHz frequencies exhibit changes from cycle to cycle. Therefore, we note that the profile seen in the folded light-curve is different from the one we would obtain in a single orbital cycle.

\subsection{Light-curve at $\bmath{\approx150}$~MHz}

Several data sets at a frequency around 150~MHz have been analyzed: one GMRT observation at 154-MHz, one LOFAR observation at 142~MHz and four LOFAR observations at 149~MHz. In the following, we will refer to all these observations generically as 150-MHz observations. We note that the differences between these frequencies would not imply any significant change in the flux density values of LS~I~+61~303 for reasonable spectral indices.

The 150-MHz GMRT data (obtained at a superorbital phase of $\phi_{\rm so} \approx 0.69$) reveal for the first time a detection of LS~I~+61~303 at this frequency\footnote{We note that LS~I~+61~303 was inadvertently detected at $\approx$150~MHz in a study of the Galactic emission made by \citet{bernardi2009} with the Westerbork Synthesis Radio Telescope. The source is marginally detected in the Figs.~2 and 3 of \citet{bernardi2009}, close to the edge of the primary beam. At these positions the flux density measurements are much less reliable, and thus we will not discuss these data here. We also note that the positions of the sources in the mentioned figures present a general displacement of about 5~arcmin to the East direction with respect to the real positions.}, being a point-like source with a flux density of $52 \pm 11~\mathrm{mJy}$.

\begin{figure}
    \includegraphics[width=0.48\textwidth]{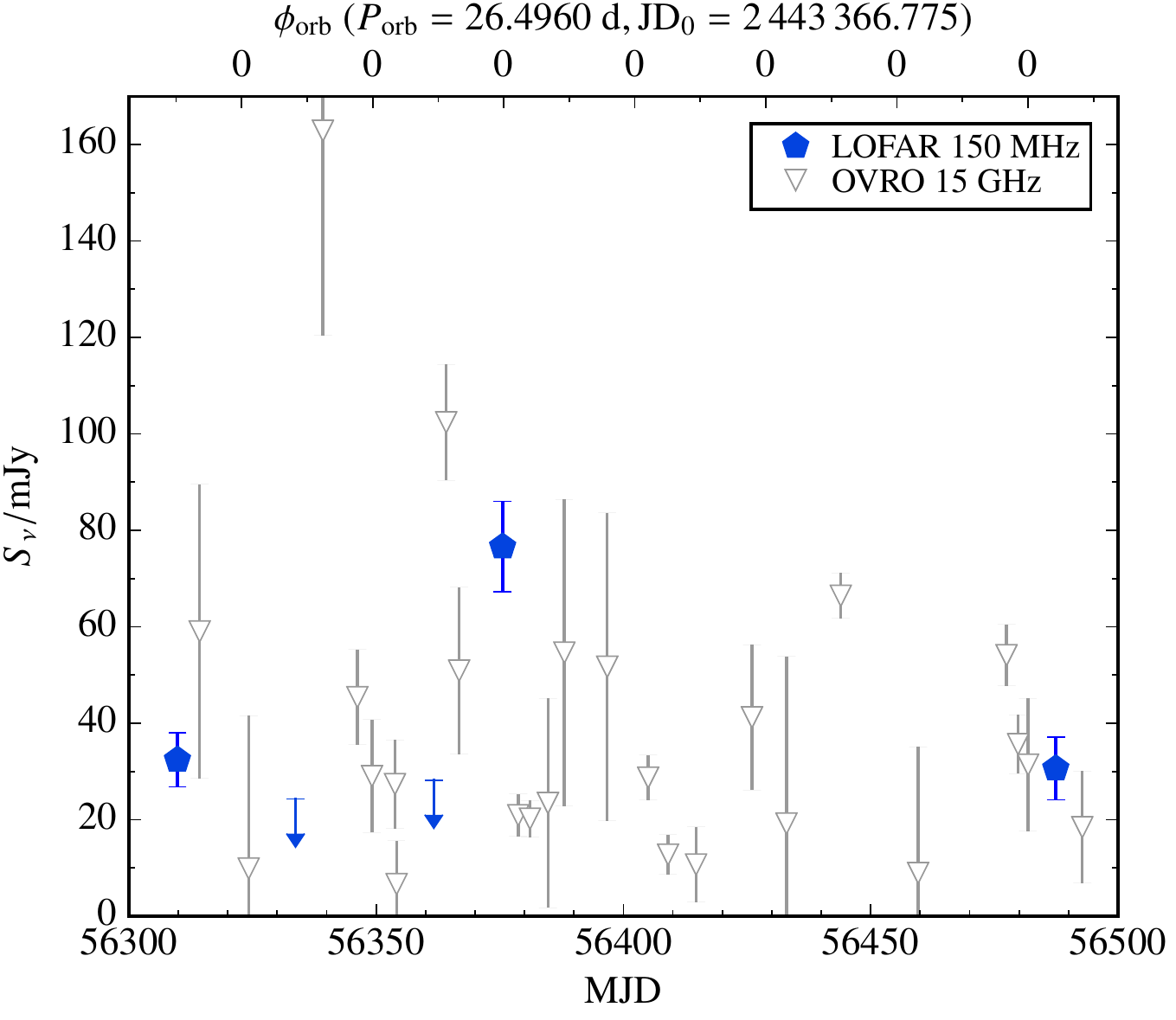}
    \caption{Flux density values of LS~I~+61~303 as a function of the MJD obtained from the 150-MHz LOFAR observations (conducted at $\phi_{\rm so} \approx 0.8$). We show the orbital phase, with labels every $\phi_{\rm orb} = 0$, on the top $x$-axis. We detect variability and a strong increase in the flux density of LS~I~+61~303 of at least a factor 3 in only 14~d (between MJD~56361 and 56375). The open triangles represent the contemporaneous OVRO data at 15~GHz.}
    \label{fig:lofar-mjd}
\end{figure}
\begin{figure}
    \includegraphics[width=0.48\textwidth]{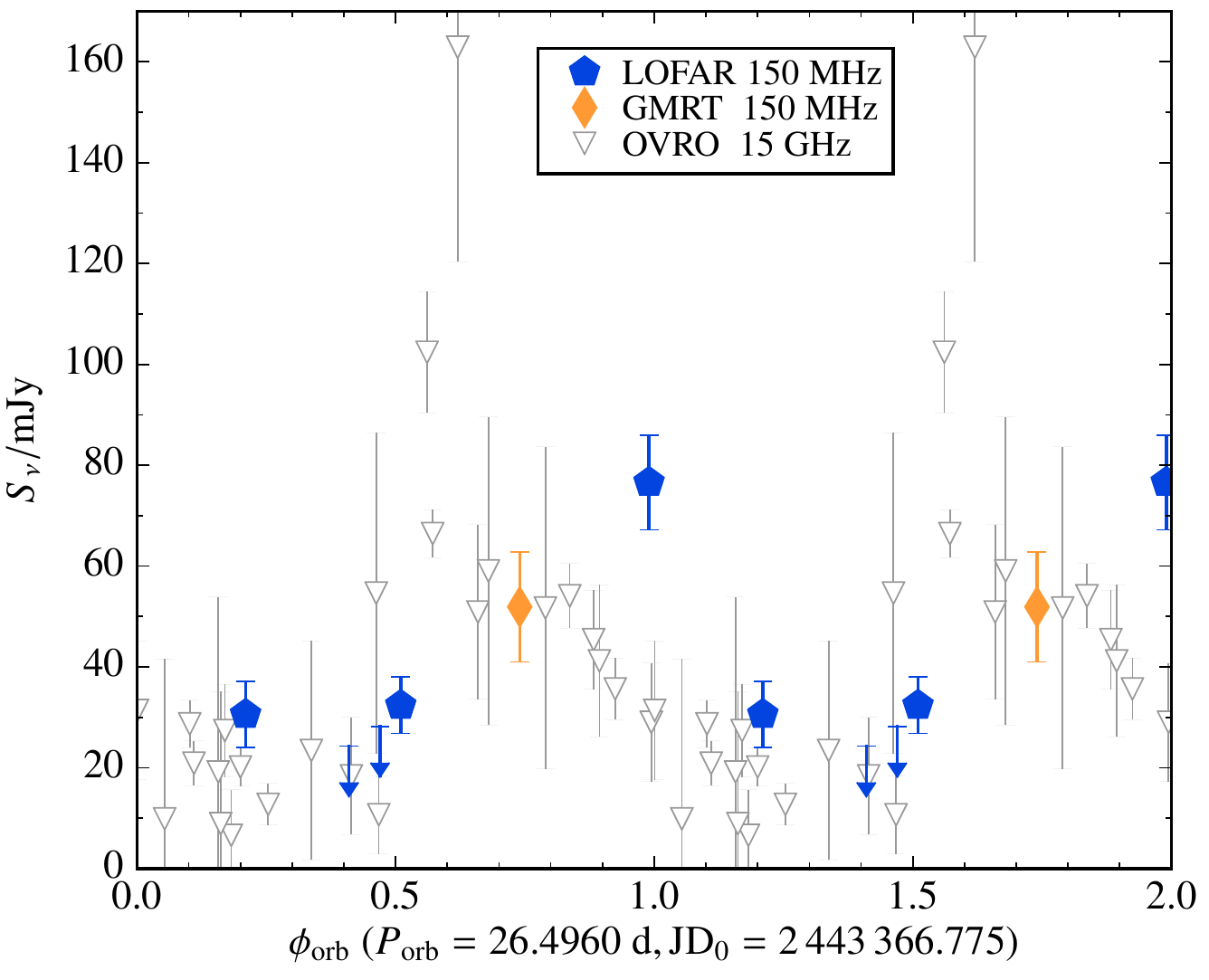}
    \caption{Folded light-curve with the orbital period from the data shown in Fig~\ref{fig:lofar-mjd} plus the 150-MHz GMRT observation (orange diamonds). The flux density is orbitally modulated, exhibiting an enhanced emission at $\phi_{\rm orb} \approx 0.7$--$1.0$. Despite the poor sampling, this maximum emission appears to be delayed $\sim$$0.3$ orbital phases with respect to the one detected in the contemporaneous OVRO data.}
    \label{fig:lofar-phase}
\end{figure}

The 150-MHz LOFAR observations taken within the RSM in 2013 ($\phi_{\rm so} \approx 0.76$--$0.87$) allow us to obtain a light-curve of the source on week timescales (spread along 7 orbital cycles). These results are shown in Fig.~\ref{fig:lofar-mjd}. With these data we clearly detect variability and a strong increase in the flux density of at least a factor 3 in only 14~d between MJD~56361 and 56375 (approximately half of the orbital period).
Folding these data with the orbital period, and adding the GMRT data at approximately the same frequency, we obtain the light-curve shown in Fig.~\ref{fig:lofar-phase}. We note that all these observations were conducted at similar superorbital phases.
We observe a kind of baseline state with flux densities around $30~\mathrm{mJy}$ on top of which there is a stronger emission between $\phi_{\rm orb} \approx 0.7$--$1.0$ (given the reduced coverage of the orbit, we certainly cannot constrain this range and it could actually be wider). Despite the poor sampling, the onset of the outburst at 150~MHz could take place at the same orbital phase as in the contemporaneous 15-GHz OVRO data, although the maximum of the outburst appears to be delayed at 150~MHz.
The average flux density for these 150-MHz observations is $35 \pm 16~\mathrm{mJy}$ (considering the upper-limits as the possible flux density value of the source: $S_{\nu} \approx 3\sigma \pm \sigma$).

\section{Discussion}\label{sec:discussion}

We have presented here data of the gamma-ray binary LS~I~+61~303 at low radio frequencies, from 150 to 610~MHz.
This is the first time that a gamma-ray binary is detected at 150~MHz, although previous searches have been performed in other sources \citep[see][for the case of LS~5039]{marcote2015ls5039}. A multifrequency monitoring conducted with the GMRT in 2005--2006 (at a superorbital phase of $\phi_{\rm so} \approx 0.2$) shows significant variability at 610~MHz with the maximum emission at orbital phases of $\phi_{\rm orb}\approx 0.8$--$1.1$. This variability is roughly coincident with the outbursts observed at 15~GHz, but with a significantly wider shape, a delay of about 0.2 orbital phases, and a slower decay. However, at 235~MHz we show that the maximum emission of the source occurs in the range between $\phi_{\rm orb} \approx 0.3$ and $\approx0.7$. The light-curve is thus almost anti-correlated with respect to the 610-MHz one. The LOFAR observations were conducted, in contrast, at a superorbital phase of $\phi_{\rm so} \approx 0.8$. In this case we observe a behaviour similar to the one observed at 610~MHz: a large variability with the maximum emission taking place at orbital phases between $0.7$ and $1.0$.

In this section we discuss the observed behaviour at 150, 235 and 610~MHz and its relationship with the superorbital modulation. It must be noted that the compact VLBI radio emission detected at GHz frequencies represents $\gtrsim$90\% of the total flux density of LS~I~+61~303, leaving small room for extended radio emission \citep{paredes1998}. For that reason, and also because of the limited number of frequencies with available light-curves (and the use of data from different orbital cycles), we assume a one-zone model with an homogeneous emitting region. We consider here the two most probable absorption mechanisms: free-free and synchrotron self-absorption (there are not enough data to consider the Razin effect such as in \citealt{marcote2015ls5039}). We infer the physical changes required in the system to explain the observed delay between the orbital phases at which the maximum emission takes place at different frequencies. 

From Fig.~\ref{fig:gmrt-mjd-all} we observe that the superorbital phase still has an important role at these low frequencies, as we detect a much larger flux density at high $\phi_{\rm so}$ than at low ones. Given that we only have two isolated observations at high superorbital phases, we cannot determine what was the state of the source during these observations (maximum of the orbital variability, minimum or in between). Therefore, we can not accurately estimate the increase of the emission as a function of the superorbital phase.

A correlation between changes in the thermal emission of the circumstellar disc, observed from optical photometry and H$\alpha$ observations, and the superorbital modulation of the non-thermal emission has been reported \citep[][and references therein]{paredes-fortuny2015}. A larger amount of material in the disc (higher values of H$\alpha$ equivalent width) is observed at superorbital phases coincident with the maximum emission at GHz frequencies ($\phi_{\rm so}\sim 0.8$--$0.0$). This larger amount of material in the disc might imply more target material for a shock with the putative pulsar wind. This could potentially lead to a more efficient particle acceleration and stronger emission, but also to a larger absorption at low radio frequencies. The presence of stronger radio emission and a clear variability at 150~MHz at these superorbital phases indicates that the emission increase, probably produced as a result of the more efficient particle acceleration, seems to dominate over the decrease due to enhanced absorption at these low radio frequencies.

The positive spectral indices obtained from the GMRT monitoring (Fig.~\ref{fig:gmrt-phase}, bottom) confirm the suggestion of a turnover between 0.3 and 1.4~GHz made by \citet{strickman1998}, and we constrain it to be in the 0.6--1.4~GHz range.
Assuming that the spectrum at low frequencies is dominated by FFA we can estimate the radius of the emitting region with the condition that the free-free opacity is
\begin{equation}
    \tau_{\rm ff} \approx 30 \dot M_{-7}^2 \nu_{\rm GHz, max}^{-2} \ell_{\rm AU}^{-3} v_{\rm W, 8.3}^{-2} T_{\rm W, 4}^{-3/2} = 1,
\label{eq:tauff}
\end{equation}
where $\dot M_{-7}$ is the mass-loss rate of the companion star in units of $10^{-7}~\mathrm{M_{\sun}\ yr^{-1}}$, $\nu_{\rm GHz, max}$ is the frequency at which the maximum emission takes place (or turnover frequency), in GHz units, $\ell_{\rm AU}$ is the radius in the spherically symmetric case, measured in AU, $v_{\rm W, 8.3}$ is the velocity of the stellar wind, in units of $2 \times 10^{8}~\mathrm{cm\ s^{-1}}$ and $T_{\rm W, 4}$ is the wind temperature, in units of $10^4~\mathrm{K}$ \citep{rybicki1979}.
Assuming a luminosity of $L \sim 4.65~\mathrm{L_{\sun}}$, typical from B0~V stars, we derive a mass-loss rate of $\dot M_{-7} \sim 0.5$ \citep{howarth1989}, for which we will assume reasonable values in the range of $\dot M_{-7} \sim 0.2$--$1$. Considering the effective temperature of $T_{\rm eff} \sim 28\,000~\mathrm{K}$, also expected for this type of star \citep{cox2000}, we deduce a wind velocity of $1\,500 \pm 500~\mathrm{km\ s^{-1}}$ (or $v_{\rm W, 8.3} \sim 0.75$, \citealt{kudritzki2000}). We also expect a wind temperature of $\sim$10\,000~K at a distance of the order of the apastron \citep{krticka2001}, and a turnover frequency $\nu_{\rm GHz, max} \sim 1$ (as mentioned before). With these values we estimate a radius for the spherically emitting region of $2.4_{-1.1}^{+1.7}~\mathrm{AU}$.

We can compare this result with the displacements of the peak positions obtained with VLBA observations along one orbital cycle by \citet{dhawan2006}.
From their Fig.~4 we observe displacements as large as $\sim$2.5 and $\sim$2~mas ($\sim$5 and $\sim$4~AU) at 2.2 and 8.4~GHz, respectively, which in our spherically symmetric model imply radii of $\sim$2.5 and $\sim$2~AU. Although we are using a spherical model, we note that the radius that we have derived is clearly compatible with these values, as expected considering that the emission has to be produced far away enough from the massive star to avoid being in the optically thick part of the spectrum due to free-free absorption.
\begin{figure}
    \includegraphics[width=0.475\textwidth]{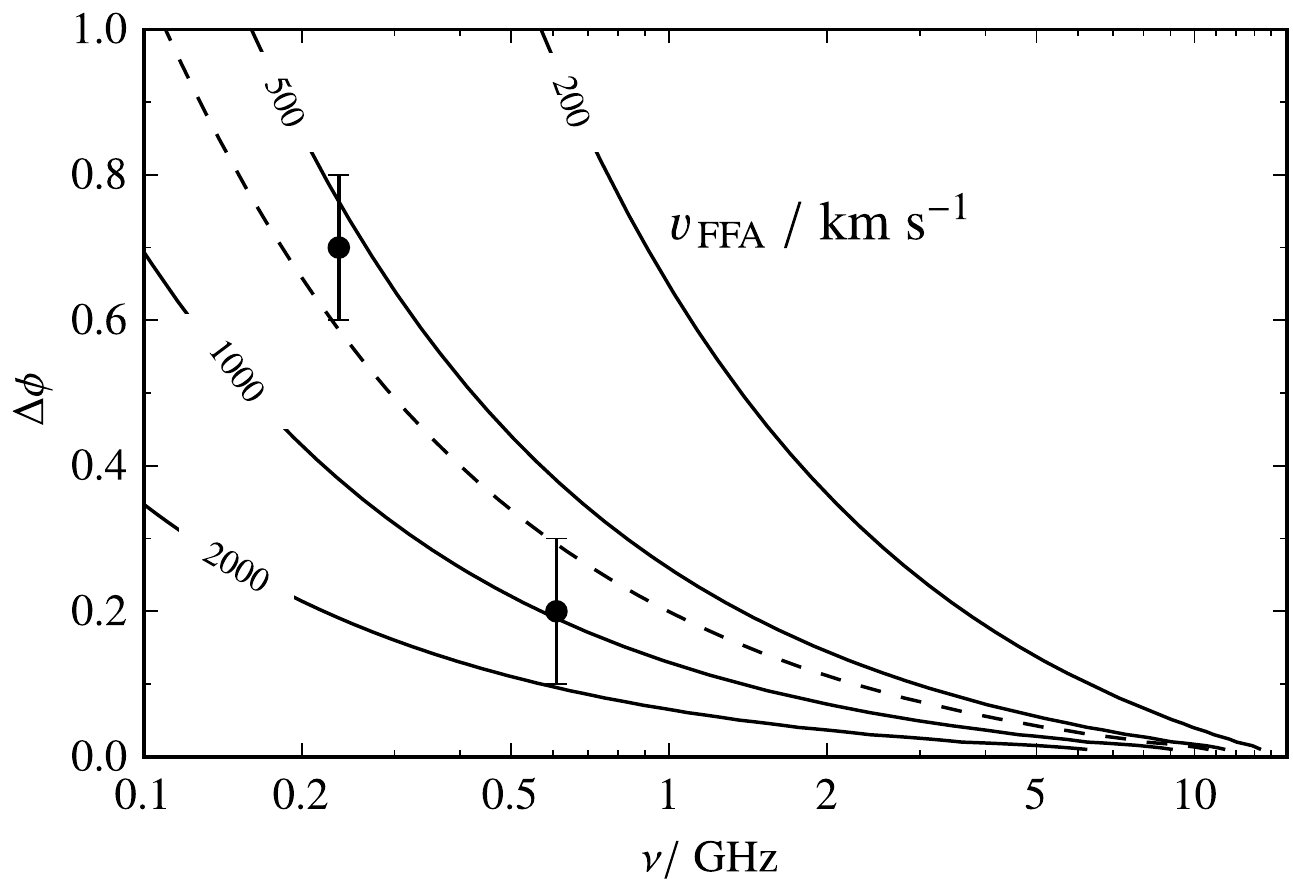}
    \caption{Shift in orbital phase ($\Delta \phi$) expected for the maxima in the flux density emission between a frequency $\nu$ and 15~GHz for different velocities of expansion of the radio emitting region ($v_{\rm FFA}$) assuming that FFA is the dominant absorption process and following equation (\ref{eq:phiff}). The black circles represent the shifts at 235 and 610~MHz with respect to 15~GHz in the folded light-curve considered in the discussion. We fit the data with an expansion velocity of $\sim$$600~\mathrm{km\ s^{-1}}$ (dashed line).}
    \label{fig:shift}
\end{figure}

\begin{figure}
    \includegraphics[width=0.475\textwidth]{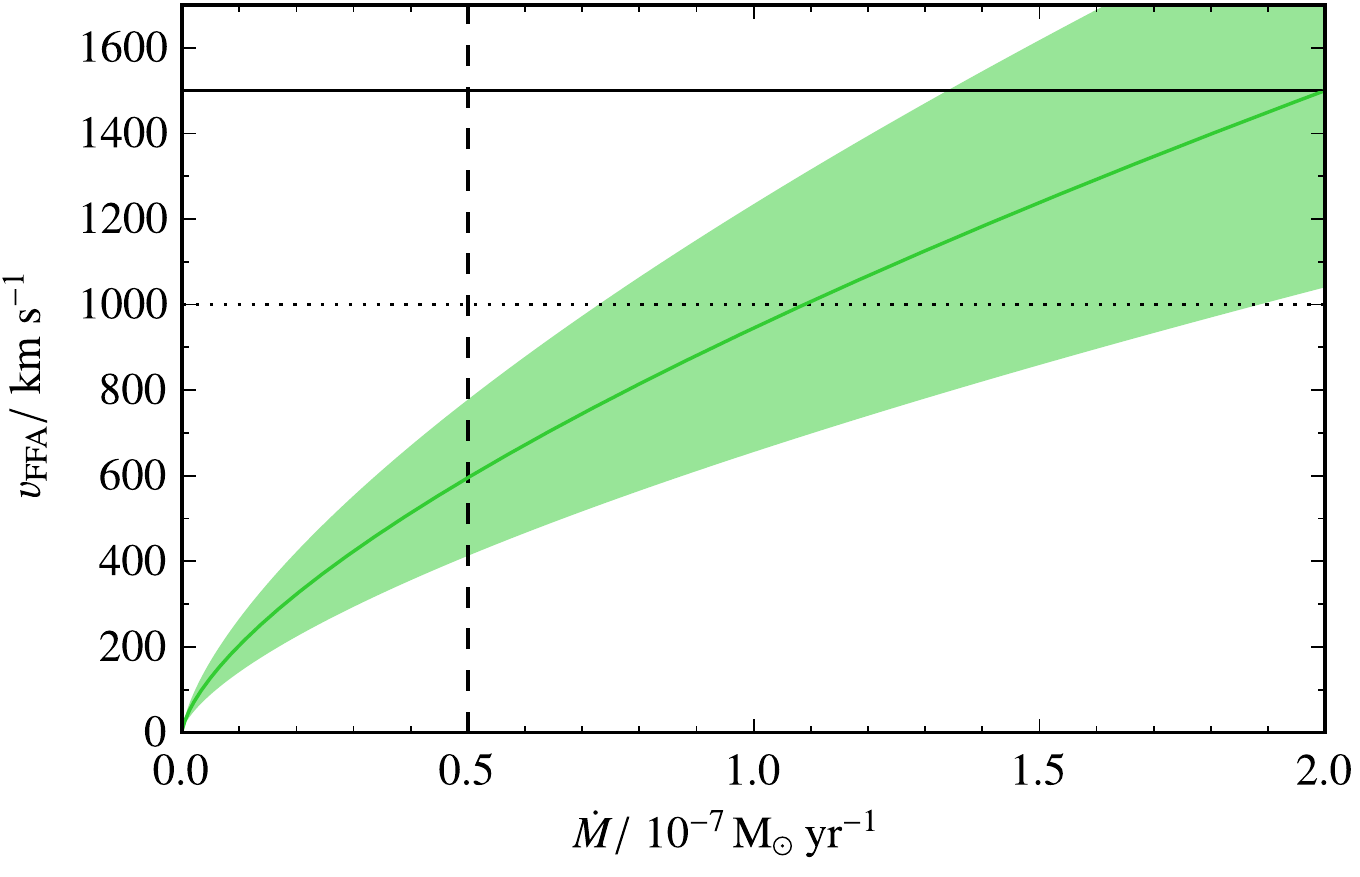}
    \caption{Velocities, $v_{\rm FFA}$, of the expanding emitting region as a function of the mass-loss rate, $\dot M$, derived from the considered delay between the maxima at 235 and 610~MHz with respect to 15~GHz, assuming a free-free absorbed region. The green curve represents the mean derived velocity and the light green shaded region represents the possible velocity values by considering the uncertainties in the orbital phase shifts between different frequencies, and in the stellar wind velocity. The vertical dashed line denotes the derived mass-loss rate. The horizontal lines represent the mean and lower value of the stellar wind velocity (solid and dotted line, respectively).}
    \label{fig:mdot-velocity}
\end{figure}

On the other hand, we can consider that the emitting region is expanding, and thus the turnover frequency evolves along the time. In this case we consider that the transition from an optically-thick to an optically-thin region would produce the delay between the different maxima observed at 235 and 610~MHz, as previously reported for other binary systems \citep{ishwarachandra2002}. From equation (\ref{eq:tauff}) we can determine the radius of the emitting region when the turnover is located at a frequency $\nu$:
\begin{equation}
    \ell_{\nu} = 30^{1/3}\, \dot M_{-7}^{2/3}\, v_{\rm W, 8.3}^{-2/3}\, T_{\rm W,4}^{-1/2}\, \nu^{-2/3}.
    \label{eq:ellff}
\end{equation}
With this relation we can infer the velocity of the expanding emitting region, or expansion velocity, assuming it to be constant for simplicity:
\begin{equation}
    v = \frac{\Delta \ell}{\Delta t} = \frac{\ell_{\nu_2}-\ell_{\nu_1}}{\Delta \phi P_{\rm orb}},
    \label{eq:vff}
\end{equation}
where $\Delta \phi$ is the shift in orbital phase for the maximum at a frequency $\nu_2$ and $\nu_1$, and $P_{\rm orb}$ is the orbital period.
We can consider for simplicity the case that $\dot M_{-7}$, $v_{\rm W, 8.3}$ and $T_{\rm W,4}$ remain constant during this expansion.
Figs.~\ref{fig:gmrt-mjd} and \ref{fig:gmrt-phase} show that the peak of the emission at 610~MHz is located somewhere between $\phi_{\rm orb} \sim 0.8$ and $1.1$. In parallel, we observe that the flux density at 235~MHz remains increasing at $\phi_{\rm orb} \sim 0.4$, and in the range of 0.6--0.7 the flux density is already decreasing. Therefore, the maximum emission probably takes place in the range of 0.3--0.7. These values imply that the maximum emission exhibits a shift of $\approx$$0.2$--$0.9$ in orbital phase between the two frequencies. Although the uncertainty in this shift is large, we can assume a shift of $\sim$0.5 to provide a rough estimation of the expansion velocity for the radio emitting region. Assuming this value and with all the previously mentioned data, we infer that the emitting region should expand by a factor of $2.0 \pm 0.5$ during this delay of $\approx$0.5 in orbital phase ($\approx$$13~\mathrm{d}$) from 610 to 235~MHz, implying a constant expansion velocity of $v_{\rm FFA} = 400_{-200}^{+300}~\mathrm{km\ s^{-1}}$.
If we also consider the shifts in orbital phase with respect to the maximum at 15~GHz,
we can deduce, from equations (\ref{eq:ellff}) and (\ref{eq:vff}), the expected delay in orbital phase as a function of the velocity, $v_{\rm FFA}$, and the frequency, $\nu$:
\begin{equation}
    \Delta\phi_{\rm FFA} =  \frac{30^{1/3}\, \dot M_{-7}^{2/3}\, v_{\rm W, 8.3}^{-2/3}\, T_{\rm W,4}^{-1/2}}{v_{\rm FFA} P_{\rm orb}} \left(\nu_{\rm GHz}^{-2/3} - 15^{-2/3} \right).
    \label{eq:phiff}
\end{equation}
Fig.~\ref{fig:shift} shows the velocity curves derived from this equation, where a fit to the considered shifts at 235 and 610~MHz with respect to 15~GHz produces a constant velocity of $v_{\rm FFA} = 600 \pm 200~\mathrm{km\ s^{-1}}$. This velocity is a factor of 2--3 smaller than the one of the stellar wind.
However, the derived velocity depends strongly on the assumed mass-loss rate, which is unconstrained from the observational point of view. Fig.~\ref{fig:mdot-velocity} shows the derived expansion velocity of the emitting region for different values of the mass-loss rate (in Fig.~\ref{fig:shift} we assumed $\dot M_{-7} = 0.5$). We note that for mass-loss rates of $\dot M_{-7} \approx 1$ (which is still possible) the derived velocities overlap with the range of possible stellar wind velocities. In any case, these velocities are much lower than the relativistic velocity of the putative pulsar wind (see discussion in \citealt{bogovalov2012}).

\citet{dhawan2006} estimated the outflow velocity along the orbit at 2.2 and 8.4~GHz, obtaining maximum values of $\sim$$7\,500~\mathrm{km\ s^{-1}}$ near periastron and $\sim$$1\,000~\mathrm{km\ s^{-1}}$ near apastron. Therefore, the expansion velocities derived here are also close to the one derived by these authors near apastron.

\begin{figure}
    \includegraphics[width=0.475\textwidth]{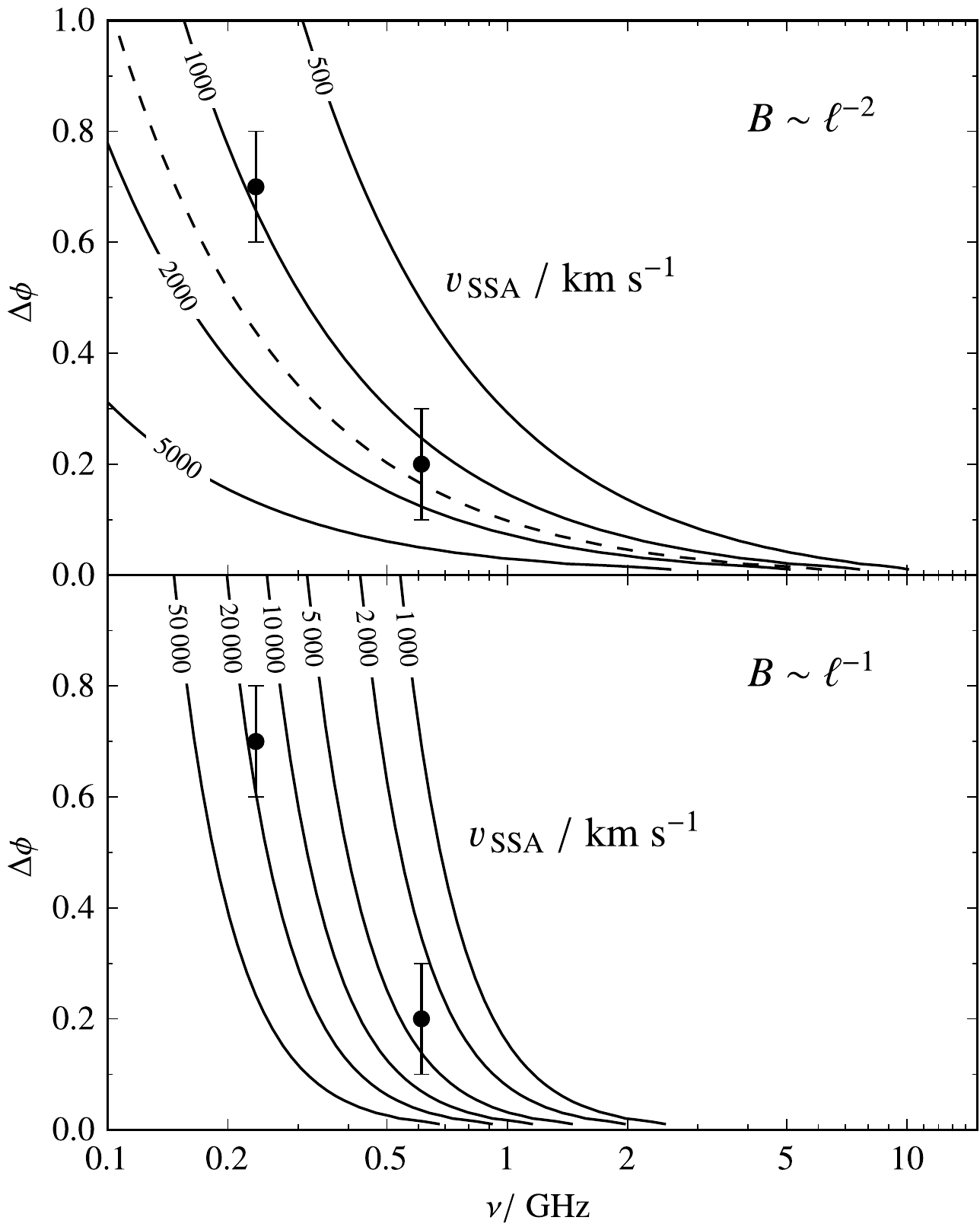}
    \caption{Same as Fig.~\ref{fig:shift} but assuming that SSA is the dominant absorption process and following equation (\ref{eq:phissa}). Two different dependences of $B$ have been considered: $B \sim \ell^{-2}$ (top) or $B \sim \ell^{-1}$ (bottom). The dashed line on the top panel represents the mean value of the derived wind velocity. We note that with the dependence $B\sim \ell^{-2}$ (top) we can easily explain the data with an expansion velocity of $\sim$$1\,000~\mathrm{km\ s^{-1}}$, while with the dependence $B \sim \ell^{-1}$ (bottom) the data can not be reproduced with a constant velocity.}
    \label{fig:shiftSSA}
\end{figure}

If instead of FFA we consider SSA as the dominant absorption process at low frequencies, we can estimate the properties of the emitting region in the optically thick to optically thin transition. In this case, the SSA opacity is
\begin{equation}
    \tau_{\rm SSA} = 3.354 \times 10^{-11} (3.54 \times 10^{18})^{p} K B^{(p+2)/2} b(p) \nu_{\rm GHz}^{-(p+4)/2} \ell_{\rm AU} = 1,
    \label{eq:taussa}
\end{equation}
where $K$ is the normalization of the accelerated particles, $B$ is the magnetic field, and $b(p)$ is a function that only depends on $p \equiv 1-2\alpha$ \citep{longair2011}.
Assuming $p \sim 2$ (i.e. $\alpha \sim -0.5$ in the optically thin region, similar to the average spectral index observed in LS~I~+61~303 above $\sim$2~GHz) we would expect that the quantity $K\,B^2\,\ell$ decreases a factor of $\sim$18 during the delay of 0.5 in orbital phase between the maximum emission at 235 and 610~MHz. 
These three quantities ($K, B, \ell$) are coupled in the equation above and thus we can not estimate them separately. All of them would a-priori change along the orbit, but since we assume an expanding region from the same population of accelerated particles, $K$ should remain constant (provided that losses are not significant). 
Additionally, one can consider different dependences of $B$ as a function of $\ell$: $B \sim \ell^{-2}$ (as in a spherical expansion) or $B \sim \ell^{-1}$ (as in a conical or toroidal expansion, as it happens either in a relativistic jet from the microquasar scenario or in a cometary tail from the young non-accreting pulsar scenario). With these considerations we expect an expansion factor of $\sim$2.6 (if $B \sim \ell^{-2}$) or $\sim$18 (if $B \sim \ell^{-1}$).
To derive the expansion velocity of the emitting region, we can assume the radius of $\ell = 2.4_{-1.1}^{+1.7}~\mathrm{AU}$ that we have obtained at 1~GHz in the FFA case, which is also compatible with the results from \citet{dhawan2006}.
With these assumptions, and the equations (\ref{eq:vff}) and (\ref{eq:taussa}), we obtain the shifts in orbital phase between the maximum at a frequency $\nu$ and at 15~GHz:
\begin{equation}
    \Delta\phi_{\rm SSA} =  \frac{\nu_{\rm GHz}^3 - 15^3}{v_{\rm SSA} P_{\rm orb} 3.354 \times 10^{-11} (3.54 \times 10^{18})^{2} K B^{2} b(2)}.
    \label{eq:phissa}
\end{equation}
Fig.~\ref{fig:shiftSSA} shows the derived velocity curves as a function of the frequency and the orbital phase shift for the SSA case. In the case of $B \sim \ell^{-2}$ we can fit the data with a constant velocity of $v_{\rm SSA} = 1\,000_{-500}^{+600}~\mathrm{km\ s^{-1}}$, where the uncertainties result from the propagation of the uncertainties in $\ell$ at 1~GHz. The fitted velocity is also compatible with the one of the stellar wind. In the case of $B \sim \ell^{-1}$ we obtain velocity curves that can not reproduce the considered phase delay between frequencies. A constant velocity of $17\,000_{\ \,-8\,000}^{+12\,000}~\mathrm{km\ s^{-1}}$ produces the best fit to the data, although with a reduced $\chi^2$ of 2.6. In any case, these velocities are significantly faster than the stellar wind velocity, but clearly slower than the relativistic velocity of the putative pulsar wind. In conclusion, a dependence of $B \sim \ell^{-2}$ is more plausible in the considered model where SSA dominates. This would suggest a different geometry than the one expected in the considered scenarios, but compatible with our initial assumption of a simple spherically symmetric emitting region. 
We note that if energy losses were considered, the derived expansion velocities would be even lower than the ones discussed.


Extrapolating these results to 150~MHz, we infer an orbital phase delay of about 0.9 (assuming FFA) or $\sim$1.0 (assuming SSA).
However, these results cannot be directly compared with the obtained light-curves at 150~MHz. These data were taken at different superorbital phases ($\phi_{\rm so} \sim 0.9$ instead of $\sim$0.2), and we observe a larger flux density emission and a larger variability at these 150~MHz data than the ones expected from an extrapolation of the 235 and 610-MHz light-curves. This implies, as mentioned before, that the superorbital phase still plays a significant role at low frequencies.
In any case, the shift at 150~MHz with respect to the 15-GHz data is not well constrained and it could be between $\sim$0.0 and $\sim$0.5 orbital phases, with the possibility of observing a full cycle shift (and thus between 1.0 and 1.5). We note that these values are roughly compatible with the ones derived from the 235 and 610-MHz data assuming a one-cycle delay. However, we cannot discard the possibility of being observing a delay of only $\sim$0.0--0.5 orbital phases, as we would expect less absorption at these high superorbital phases, which would also explain why we see a large orbital variability.


\section{Conclusions}\label{sec:conclusions}

We have detected for first time a gamma-ray binary, LS~I~+61~303, at a frequency as low as $150~\mathrm{MHz}$. This detection establishes the starting point to explore the behaviour of gamma-ray binaries in the low frequency radio band, which will allow us to unveil the absorption processes that can occur in their radio spectra or light-curves.
Additionally, we have obtained light-curves of LS~I~+61~303 at 150, 235 and 610~MHz, observing orbital and superorbital variability in all cases. In the folded light-curve with the orbital period we observe quasi-sinusoidal modulations with the maxima at different orbital phases as a function of the frequency. The observed delays between frequencies seem to be also modulated by the superorbital phase. The flux density values are also modulated by the superorbital phase, with the source displaying a stronger emission at $\phi_{\rm so} \sim 1$.

Assuming a simple spherically symmetric model we have obtained a coherent explanation for the delays considered between the maxima as a function of the frequency. Within this model, we observe that either FFA or SSA can explain the delays of the low-frequency emission of LS~I~+61~303. We have also estimated an expansion velocity for the radio emitting region of $\sim$$1\,000\ \mathrm{km\ s^{-1}}$ for both absorption mechanisms, which is close to the stellar wind velocity. The obtained velocity could hardly be obtained if there was a relativistic jet, expected in the microquasar scenario, giving further support to the scenario involving a shock between the relativistic wind of a young non-accreting pulsar and the non-relativistic stellar wind.
In the case of SSA, a decay of $B \sim \ell^{-2}$ (as in a spherical expansion) is supported. This is compatible with the geometry of our model, but unexpected considering the possible geometries in the different scenarios.

The limited amount of data acquired up to now precludes detailed modelling to establish the origin of the variability at different frequencies and epochs.
Further multi-epoch observations of LS~I~+61~303 with LOFAR are needed. These data would allow us to determine the light-curve of LS~I~+61~303 folded in orbital phase and study its changes as a function of the superorbital phase. A good coverage of a single orbital cycle is mandatory to obtain a reliable profile of the variability of LS~I~+61~303 due to the significant differences observed between outbursts at different orbital phases.
Future simultaneous multifrequency observations with the GMRT, LOFAR and the VLA at different superorbital phases would allow us to study the light-curve of LS~I~+61~303 and its dependence with the frequency. These data could unveil the changes in the physical parameters that characterise the emitting region and the absorption processes required to explain the superorbital modulation.
Finally, the use of the International stations in LOFAR observations (longer baselines) would allow us to search for the extended emission at arcsec scales that is expected to arise at low frequencies.

\section*{Acknowledgments}

We thank the anonymous reviewer for providing comments that helped to improve the original version of the manuscript.
We thank the staff of the GMRT and LOFAR who made these observations possible.
GMRT is run by the National Centre for Radio Astrophysics of the Tata Institute of Fundamental Research.
LOFAR, the Low Frequency Array designed and constructed by ASTRON, has facilities in several countries, that are owned by various parties (each with their own funding sources), and that are collectively operated by the International LOFAR Telescope (ILT) foundation under a joint scientific policy.
The Ryle Telescope is operated by the University of Cambridge and supported by STFC.
This research has made use of data from the OVRO 40-m monitoring program which is supported in part by NASA grants NNX08AW31G and NNX11A043G, and NSF grants AST-0808050 and AST-1109911. We thank T. Hovatta for providing the OVRO 40-m dish data.
B.M., M.R. and J.M.P. acknowledge support by the Spanish Ministerio de Econom\'ia y Competitividad (MINECO) under grants AYA2013-47447-C3-1-P, FPA2013-48381-C6-6-P, MDM-2014-0369 of ICCUB (Unidad de Excelencia `Mar\'ia de Maeztu'), and the Catalan DEC grant 2014 SGR 86.
B.M. acknowledges financial support from MINECO under grant BES-2011-049886.
J.M.P. acknowledges financial support from ICREA Academia. J.W.B. acknowledges support from European Research Council Advanced Grant 267697 `4 Pi Sky: Extreme Astrophysics with Revolutionary Radio Telescopes'. R.P.B. has received funding from the European Union Seventh Framework Programme under grant agreement PIIF-GA-2012-332393. S.C. acknowledges financial support from the UnivEarthS Labex program of
Sorbonne Paris Cit\'e (ANR-10-LABX-0023 and ANR-11-IDEX-0005-02). The financial assistance of the South African SKA Project (SKA SA) towards this research is hereby acknowledged. Opinions expressed and conclusions arrived at are those of the authors and are not necessarily to be attributed to the SKA SA.

\bibliographystyle{mn2e}
\bibliography{bibliography.bib}

\label{lastpage}

\end{document}